\documentclass[fleqn,usenatbib]{mnras}

\usepackage{newtxtext,newtxmath}

\usepackage[T1]{fontenc}

\DeclareRobustCommand{\VAN}[3]{#2}
\let\VANthebibliography\thebibliography
\def\thebibliography{\DeclareRobustCommand{\VAN}[3]{##3}\VANthebibliography}

\usepackage{multirow}

\usepackage{graphicx}	
\usepackage{amsmath}	

\renewcommand{\d}{\mathrm d}
\newcommand{\eref}[1]{Eq.~\eqref{#1}}
\newcommand\codename{\texttt{RaMiCES}}
\newcommand\modelname{\texttt{SACEM}}
\renewcommand{\d}{\mathrm d}

\defcitealias{Sneden2008}{Sn08}
\defcitealias{Siegel2019}{Si19}
\newcommand{\sneden}{\citetalias{Sneden2008}}
\newcommand{\siegel}{\citetalias{Siegel2019}}
\defcitealias{Schonrich2009}{SB09}
\newcommand{\sbRef}{\citetalias{Schonrich2009}}

\newcommand{\tc}{\tau_\text{coll}}
\newcommand{\Zc}{Z_\text{c}}
\newcommand{\Z}{\mathcal{Z}}
\newcommand\rsfr{\rho_\text{sfr}}
\newcommand\fittedImage[1]
{
	\begin{center}
 \includegraphics[keepaspectratio=true, width=0.47\textwidth]{Plots/#1}
	\end{center}
}

\newcommand\fittedImageLarge[1]
{
	\begin{center}
 \includegraphics[keepaspectratio=true, width=0.97\textwidth]{Plots/#1}
	\end{center}
}

\def\commentmode{1}
\newcommand\comment[2]
{
	\if\commentmode1
		\textcolor{red}{\footnotesize \bf (#1) #2}
	\fi
}

\title[Collapsars Cannot be Dominant r-Process Source in MW]{Metallicity-Suppressed Collapsars Cannot be the Dominant r-Process Source in the Milky Way}

\author[J. Fraser \& R. Sch{\"o}nrich]{
Jack Fraser,$^{1}$\thanks{E-mail: jackthomas.fraser@physics.ox.ac.uk}
Ralph Sch{\"o}nrich,$^{2}$
\\
$^{1}$Rudolf Peierls Centre for Theoretical Physics, Clarendon Laboratory, Parks Road, Oxford OX1 3PU, UK\\
$^{2}$Mullard Space Science Laboratory, University College London, Holmbury St Mary, Dorking, Surrey RH5 6NT, UK\\
}

\begin{document}

\label{firstpage}
\pagerange{\pageref{firstpage}--\pageref{lastpage}}
\maketitle

	\begin{abstract}
	
	 We develop a high-performance analytical model of Galactic Chemical Evolution, which accounts for delay time distributions and lock-up of stellar yields in a thermal-phased ISM. The model is capable of searching, for the first time, through the high-dimensional parameter space associated with the r-process enrichment of the Milky Way by its possible sources: Neutron Star Mergers and Collapsar events. Their differing formation mechanisms give these two processes different time dependencies, a property which has frequently been used to argue in favour of collapsars as the dominant r-process source. However, we show that even with large degrees of freedom in the allowed thermal, structural, and chemical properties of the galaxy, large regions of parameter space are in strong tension with the data. In particular, whilst we are able to find models in which neutron star mergers produce the majority of r-process material, the data rule out all models with dominant collapsar yields. With no other identified source, we conclude that Neutron Star Mergers must be the dominant contributors to the modern Milky Way r-process budget.

	
	\end{abstract}

\section{Introduction}
	
	Since the landmark paper of \cite{Burbidge1957}, it has been widely accepted that, in order to explain the abundance distribution of chemical elements observed in the universe, we require a number of distinct nucleosynthesis channels, operating in unison.

	Primordial nucleosynthesis sourced the lightest elements in the universe (H and He, \citealt{Alpher1948}), while the heavier elements are created in processes such as shell burning in the stellar interior (the $\alpha$ elements, including O, Si and Mg, \citealt{Hoyle1954}), explosive nucleosynthesis during supernovae events ($\alpha$ elements, plus the iron peak elements: Fe, Ni, Co, \citealt{Arnett1970}) or cosmic ray spallation (Li, Be, B,  \citealt{Reeves1970}).

	The majority of the elements heavier than iron, however, are sourced from a variety of neutron-capture processes: the slow (s), intermediate (i) and rapid (r) neutron capture processes. Whilst the origin of the \textit{s}-process is well understood \citep{Clayton1961}, and the \textit{i}-process thought to contribute significantly to only a handful of isotopes \citep{Cote2018}, the debate about the astrophysical sites that can lead to sufficient r-process synthesis has remained an open and enduring question for many years, with several possible sites for the r-process being identified:

	\cite{Woosley1994} argue that neutrino heating in Core Collapse Supernovae (CCSN) creates favourable conditions such that normal CCSN can provide a site for the r-process. This model, however, is plagued by {overproduction} of certain elements, and the high entropy conditions required have been questioned in more recent studies (i.e. \citealt{Fischer2012}). In addition, the existence of ultra metal poor, but highly r-process enriched stars (such as that found in \citealt{Sneden1996}) indicates that the source of r-process nucleosynthesis must be a rare, high yield event. CCSN are therefore not considered a viable source of r-process enrichment.
	
	The disruption of a neutron star by tidal interactions during a merger with a black hole \citep{Lattimer1974}, or by a binary collision between two neutron stars \citep{Symbalisty1982,Freiburghaus1999,Rosswog1999} are also candidates for the r-process. The detection of the combined gravitational wave GW170817 and GRB event GRB170817A, confirmed to arise from a NS-NS merger event \citep{Abbott2017}, and the subsequent detection of r-process material in the ejecta \citep{Chornock2017,Rosswog2018} provided direct observational evidence that Neutron Star Mergers (NSM) produce r-process material.
	
	Though the existence of NSM as an active r-process pathway is rarely called into question, it is seen as concerning that time-delayed nature of NSM formation would na\"ively predict entirely different enrichment pathways in [Eu/Fe]-[Fe/H] space than is observed, leading to either the conclusion that NSM cannot be dominant r-process sources \citep{Argast2004, Wanajo2006}, or the invocation of neutron star properties incompatible with their understood behaviour \citep{Matteucci2014}.
	
	In an alternative approach, \cite{Fujimoto2006} combined an MHD jet method with the collapsar models of \cite{Woosley1993}, and demonstrated that this can produce a significant amount of r-process enrichment. Collapsars occur when the core of a collapsing star is rotating sufficiently fast to delay radial infall, resulting in MHD jets driven by accretion onto a compact engine, and are thought to be the source of Long Gamma Ray Bursts. Although LGRB are well-documented events and often tied to unusual forms of CCSN due to their formation in regions of rapid star formation \citep{Bloom2002} and several closely tied observations of supernovae associated with LGRBs \citep{Kulkarni1998,Mazzali2003,Sollerman2006}, there is no direct evidence linking their formation with r-process synthesis. However, the model is widely favoured, since the high progenitor masses imply a short lifetime and thus allow for very early r-process enrichment.

	Other sources for r-process material have also been studied. For example, neutrino-driven winds \citep{Wanajo2001} and electron-capture supernovae \citep{Wanajo2011}, however \cite{Haynes2018} showed that these did not produce sufficient quantities of r-process material. Whilst in the case of the intermediary (\textit{i}) process-producing White Dwarf binaries \citep{Cowan1977,Denissenkov2017}, it was predicted by \cite{Cote2018} that only specific isotopes are produced in significant quantities, with 45 per cent of solar Mo predicted to be i-process in origin, but less than $10^{-2} per cent$ of solar Eu. For the sake of clarity and simplicity, we will therefore neglect these sources. 
	
	Confusing matters further, there is also evidence of an {\it incomplete} (or {\it weak}) r-process \citep{Honda2006}, in which the lighter r-process elements are synthesised, but not the heavier second and third peak elements. Magneto-rotational supernova discussed in \cite{Nishimura2017} (also referred to as `hypernovae', though this phenomenological term can refer to collapsars) or the recently proposed Quark-Deconfinement Supernovae of \cite{Fischer2020} are thought to be good candidates. For our purposes, we use the term `r-process synthesis' to refer to the {\it complete} r-process, in which all r-process material up to the third peak is synthesised.	
	
	It might feel natural to assume that multiple pathways actively contribute to the r-process enrichment, with collapsars providing the early time yield, and then NSM coming in later.  However, this bears two problems: i) the enrichment profiles for [Eu/Fe] are poorly replicated in simple GCE models whenever NSM are significant contributors, and ii) \cite{Sneden1996} demonstrated a common r-process fingerprint: a remarkable consistency of relative r-process abundances, with \cite{Sneden2008} (henceforth \sneden{}) extending this relationship. The relative abundances of r-process material in the metal-poor but highly r-process enriched star CS 22892-052 match those of the solar system, despite the fact that the high enrichment indicates very early enrichment from the unmixed ejecta of a single r-process event. The common `fingerprint' with the solar system implies that the material is produced in the same ratio throughout galactic history that dominates today's abundances, a tension if one would like to assume that the dominant\footnote{Throughout this paper we use the nomenclature that a {\it dominant} source is one which can be assumed to be the sole source, neglecting all others. A non-dominant source which produces more than 50 per cent of production is a {\it majority} source.} channel switched from collapsars early on to NSMs today. 
	
	In this paper, we will put quantitative limits on the relative contributions from both neutron star mergers and collapsars by comparing chemical evolution models with observed stellar abundances. This will show that under reasonable assumptions the relative contribution from collapsars is highly limited.
	
	Section \ref{S:ProgenitorTheory} will introduce some of the key aspects of collapsar formation as it pertains to chemical evolution, with section \ref{S:RProcessFeatures} detailing the observational data that we will attempt to replicate in our models. Section \ref{S:Models} introduces the analytical \modelname{} model, and briefly discusses the full evolutionary simulation \codename{}, whilst Section \ref{S:GridSearch} outlines our attempt to eliminate regions parameter space, and \S\ref{S:Results} and \S\ref{S:Properties} discuss our findings regarding the excluded regions of parameter space, and the required properties of our models.

\section{Modelling Collapsars \& Their Yields}\label{S:ProgenitorTheory}\label{S:Collapsar}
	
	Collapsars are a corollary to the existence of the `failed supernovae' proposed in \cite{Bodenheimer1983} and since observed by \cite{Adams2017}. `Failed Supernovae' are the fate of stars which are so massive that the usual supernova mechanism is insufficient to prevent runaway radial collapse. However, progenitors with large angular momentum cannot collapse spherically. Instead, they collapse into a compact accretion disc around the growing black hole: a potential site for r-process nucleosynthesis and LGRBs.
	
	\subsection{Dependence on Initial Metallicity}\label{S:Cutoff}
	
		The nature of collapsars necessitate a mass and core angular momentum cutoff for their formation: the star must be massive enough to defy normal supernova mechanisms, but have enough internal angular momentum to stave off radial collapse. There are also strong indications of a metallicity dependence on collapsar rates, both from models of how metallicity impacts the stellar interior (through variations in the opacity and angular momentum transport efficiencies), and from observational LGRB counts such as \cite{Perley2016}. 
		
		Theoretical models of stellar evolution quote a strict cutoff in metallicity around $\Zc = 0.1 Z_\odot$ \citep{MacFadyen1999,Yoon2005,Langer2006,Woosley2006}, up to $\Zc =0.3 Z_\odot$ \citep{Yoon2006}. In these models, stars with $Z > \Zc$ undergo core-braking to below the critical threshold, and hence cannot be progenitors of collapsar events.
		
		In contrast, whilst observational constraints of GRB events points to find suppressed collapsar activity at higher $Z$, they instead find a much larger cutoff value, as high as $Z = Z_\odot$ \citep{Wolf2007}, though they do find suppression beginning beforehand. \cite{Perley2016} propose that this discrepancy arises because there are both single-progenitor and double-progenitor pathways for GRB events, with the single progenitor pathway (isolated collapsars) dying away fastest. 
		
		Some proposed double-progenitor LGRB models (i.e. \citealt{Podsiadlowski2010}), however, source their energy from explosive detonation inside a common envelope rather than accretion disc/central engine interactions, making a significant contribution to r-process nucleosynthesis questionable. In addition, some observed LGRB events have no accompanying supernova detected \citep{Fynbo2006}, further indicating a diversity of origins for LGRBs beyond r-process-producing collapsars.
		
		In summary, current observational evidence indicates that r-process producing collapsar events form a non-constant subset of all LGRB events. This is concerning as many studies (i.e. \citealt{Siegel2019}, henceforth \siegel{}) assume that the r-process nucleosynthesis tracks the GRB rate perfectly: such models thus might source r-process elements from collapsars long after they have stopped contributing in reality.
	
	\subsection{r-Process Yields} \label{S:CollYieldBasics}
	
		In principle, one may therefore continue to search for conditions which would lead to collapsar formation, and couple these with rotational, metallicity and stellar-mass distribution functions and yield tables to simulate the `true' collapsar contribution. However, this approach is unfeasible for a number of reasons:
		\begin{itemize}
			\item The models of i.e. \cite{Heger2000} and \cite{Limongi2018} show a complex/non-linear relationship between initial rotation and final core angular momentum, barring simple predictions for the likelihood of a star going collapsar from initial conditions.
			\item The physical processes occuring in the interior of massive stars are highly inaccessible. As a result, exact numerics of the yields from a given system must be considered poorly unconstrained.
		\end{itemize}
		Without rigorous observational or theoretical constraints, our models will therefore need to utilise an approximation for $Y_\text{coll}(M,Z,v_\text{zams})$. We choose the simplest approximation: a constant yield which is suppressed at a metallicity $Z = \Zc$ (Eq.\ref{E:Suppressor}). We justify this approach in \S\ref{S:YieldApprox} and conclude that it has neglgible impact on the strength of our conclusions.

\section{Patterns in the R Process Abundances}\label{S:RProcessFeatures}
	
	In this section we delineate the main features of the r-process abundances, both in the [Eu/Fe]-[Fe/H] plane, and as a function of time. The use of Eu as a proxy for the total r-process enrichment is justified in Appendix \ref{A:Tracer}.
	
	\subsection{The Abundance Planes}
	
		As stellar ages are hard to obtain accurately, we follow \cite{Tinsley1979} by studying the chemical history of the Galaxy in the [X/Fe]-[Fe/H] plane, where:
		\begin{equation}
			\left[ \frac{\text{X}}{\text{Y}} \right] = \log_{10} \left( \frac{n_\text{X}}{n_\text{Y}} \right) - \log_{10} \left( \frac{n_\text{X}^\odot}{n^\odot_\text{Y}} \right)
		\end{equation}
		$n_\text{X}$ is the number density of species X, and $\odot$ denotes solar values. 
		
		\begin{figure}
		    \centering
			\fittedImage{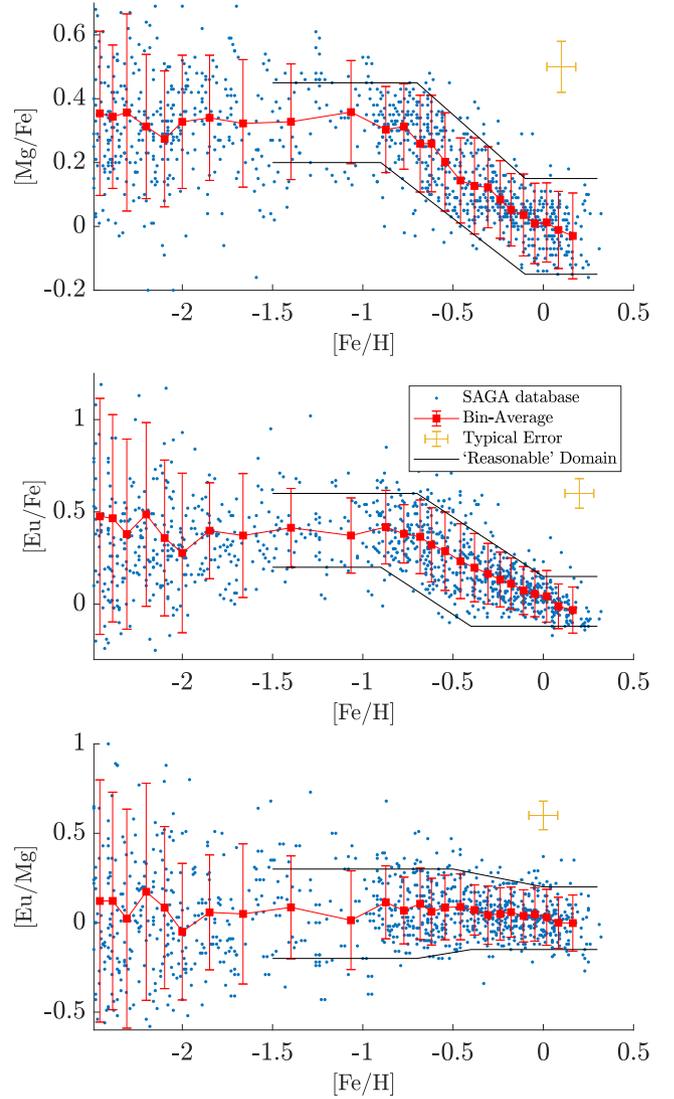}
		    \caption{Chemical abundance plots for the 965 stars in the SAGA database with Eu, Fe and Mg measurements with both upper and lower error bounds, and [Fe/H]>-2.5. In both panels, blue dots show datapoints from this SAGA subset, and red squares show the average of bins containing 40 stars,with the error bars showing the intrinsic scatter of the bin. {\it Upper Panel}: the [Mg/Fe]-[Fe/H] plane, {\it Mid Panel}: the [Eu/Fe]-[Fe/H] plane, {\it Lower Panel}: the [Eu/Mg]-[Fe/H] plane. The black lines denote the `limiting model domains', the regions which a theoretical model cannot leave for it to be considered consistent with this data, as discussed in \S\ref{S:Success}.
		}
		    \label{F:SAGAData}
		\end{figure}
		
		Figure \ref{F:SAGAData} shows a sample of 965 stars drawn from the SAGA database\footnote{\url{http://sagadatabase.jp}} which possess both upper and lower bounds for all of the elements of interest, and have [Fe/H] > -2.5, since the low metallicity end of the distribution is dominated by stochastic processes, and by stars in the galactic halo which formed within dwarf galaxies accreted during the growth of the Milky Way. The low metallicity end therefore likely represents a superposition of the chemical histories of these dwarf galaxies \citep{Ojima2018}, rather than the in-situ history of the Milky Way. Due to the low mass of such accreted objects, their final impact on galactic chemistry is negligible, and this is therefore beyond the scope of this paper to discuss.
		
		The chemical data in Fig. \ref{F:SAGAData} are sourced from a variety of surveys, and encompasses studies of disc FGK dwarfs \citep{Mishenina2013, Reddy2006}, dwarfs in both the disc and the halo \citep{Fulbright2000}, as well as studies of both giants and dwarfs in the galactic halo \citep{Sakari2018,Allen2012}, thereby providing us with a wide sample of the enrichment of the galaxy both in physical and temporal space. Following the prescriptions of \cite{Bergemann2017} and \cite{Zhao2016}, we performed a minor NLTE correction for both Mg and Eu for stars with [Fe/H] < -2, though this affected our mean trends by less than 0.02 dex, and so did not meaningfully alter our results. 
		
		The mean behaviour of [Eu/Fe], [Mg/Fe] and [Eu/Mg] for this sample is shown in red in Fig. \ref{F:SAGAData}. At low metallicity ([Fe/H] $ < -1$), we see that the bin-average shows behaviour consistent with an approximately flat curve in all three planes with [Eu/Fe] $\sim 0.4$, [Mg/Fe]$\sim0.3$ and [Eu/Mg] $\sim 0.1$. Those unfamiliar with the standard GCE models for the behaviour exhibited in Fig. \ref{F:SAGAData} may find the discussion in Appendix \ref{A:DataExplain} helpful.

	\subsection{Time Dependence}\label{S:TimeData}
	
		\begin{figure}
		    \centering
		    \fittedImage{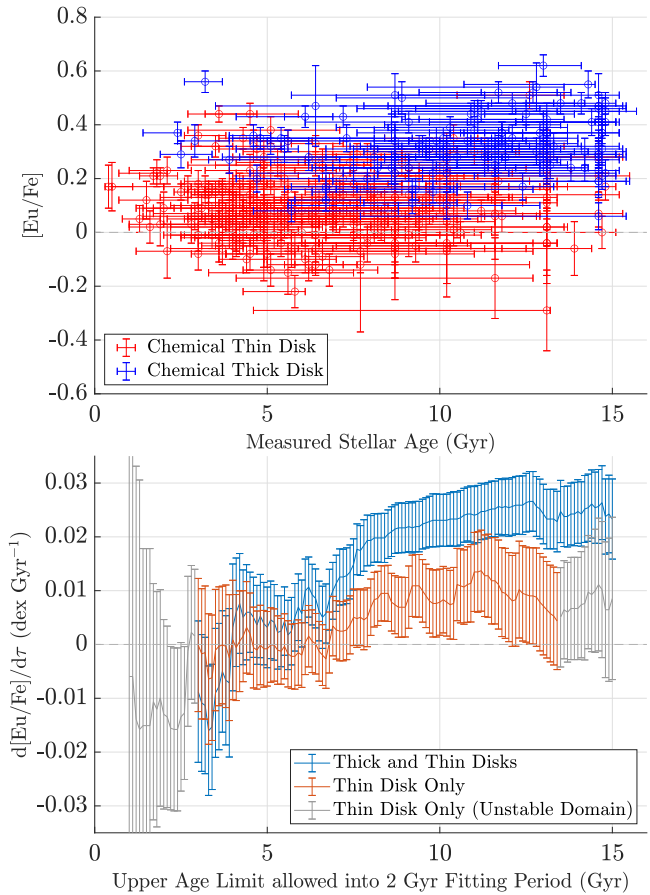}
		    \caption{Temporal abundance data derived from \protect\cite{Bensby2014} and \protect\cite{Battistini2016}. {\it Top:} The compiled data showing the inferred stellar ages and [Eu/Fe] abundances. {\it Bottom:} a plot of the calculated abundance-age gradients of each temporal subsample as a function of the maximum age permitted into the set. Bins shaded grey have fewer than 10 stars in them and so are liable to large errors.}
		    \label{F:TemporalData}
		\end{figure}
		
		We leverage the data from \cite{Bensby2014} and \cite{Battistini2016} to constrain the time evolution of the galactic chemistry, in particular the relative change of [X/Fe] with stellar age. These papers provide age and abundance estimates for 714 G and F dwarfs in the Solar Neighbourhood. A subset of 339 stars has ages and abundances for [Eu/Fe] with both upper and lower limits. These datapoints are plotted in the top panel of Fig. \ref{F:TemporalData}.
		
		The dataset also assigns membership probabilities of each star to the Galactic thick and thin discs based on kinematics only. However, due to the large overlap of disc components in kinematics, we instead prefer a chemical cut in the [Mg/Fe]-[Fe/H] plane, while we also tested that our conclusions remain unchanged irrespective of this strategy.
		
		We used a standard Bayesian approach to estimate the best-fit gradient of a given sample of stars, marginalising over unknown errors (the dataset provided only [Fe/H] uncertainties, not [Eu/H]). Using 2-Gyr long sampling periods, we build up a picture of how this gradient changes with the age of the stars, shown in the lower panel of  Fig \ref{F:TemporalData}.
		
		We see that at early times the time derivative $\d$[Eu/Fe]$/\d \tau$ strongly positive - around +0.02 dex per Gyr for the combined sample and +0.01 dex per Gyr for the chemical thin disc. A positive gradient with stellar age is equal to a negative gradient with respect to forward time, so this shows that the [Eu/Fe] abundance was decreasing during this period. 
		
		Between 7Gyr and the present day, however, gradient has decreased such that the average change in [Eu/Fe] is consistent with zero across this time period, indicating that chemical equilibrium was reached approximately 7 Gyr ago.  
		
		We note that the data of Fig. \ref{F:TemporalData} is derived under the assumption of LTE, and so both the age and abundance estimates may change under a full NLTE treatment. However, we note from Fig. 15 of \cite{Zhao2016} that significant NLTE corrections for the elements in question occur (if ever) for stars with [Fe/H]$\lesssim-1$, which implies that stars younger than 10Gyr are safe against this bias. Our conclusions about the recent equilibrium of the galaxy are therefore robust against the LTE approximation.
		
	\subsection{Star Formation Rate}

		Observational studies show that the Milky Way has sustained a Star Formation Rate until recent times which is no less than an order of magnitude below its maximal value. Though the exact value of the MW star formation rate is contested, the general consensus is that it is in the region $1 - 2  M_\odot$ yr$^{-1}$ \citep{Mor2019, Chomiuk2011, Robitaille2010, Aumer2009}.

	\subsection{Observational Summary}\label{S:ObservationalSummary}
	
		Usually models are fit to the data with the goal of optimizing to a best-fit parameter set, which is then prone to systematic biases. In this work, we walk a different path, where we instead try to falsify classes of models, based on their ability to reproduce a minimum set of constraints, which we draw from the observational evidence.
		
		For each constraint, we present a broad conclusion which can be drawn from the data (in bold), followed by how this would be replicated within a model (in italics):
		
		\newcommand\point[2]
		{
			\item \parbox[t]{7.5cm}{{#1}}
		}
		\begin{enumerate}
			\point{$[$Eu/Fe$]$ reaches $\sim0.4$ dex at [Fe/H] $\sim-1$.}{The average cannot exceed 0.5 dex or fall below 0.2 dex}
			\point{$[$Mg/Fe] reaches $\sim0.35$ dex at [Fe/H] $\sim-1$.}{The average cannot exceed 0.45 dex, or fall below 0.2 dex}
			\point{$[$Eu/Mg] is constant throughout.}{Despite a large scatter, the average remains $\sim 0.1$ dex}
			\point{Galactic chemistry is (almost) in equlibrium}{After 7 Gyrs, deviations from chemical equilibrium must be small.}
			\point{Star formation has continued until late times}{A model must maintain enough cold gas for star formation}
		\end{enumerate}

\section{Models}\label{S:Models}

	In all our models we make the simplifying approximation that there are only 3 sources of nucleosynthesis for r-process elements: a small contribution from Core Collapse Supernovae\footnote{We use CCSN as a catchall term for all processes for which the yield rate is strongly correlated with the current SFR at all points in history, and hence, via our calibration procedure (\S\ref{S:Calibrate}), includes secondary sources such as magneto-rotational supernovae (`hypernovae').}, Neutron Star Mergers, and Collapsars. Whilst other sources of r-process may exist and be important for explaining the abundance of individual stars, we expect the majority of such sources to be of subdominant importance in the case of the overall r-process trends in the galaxy. In the case of BH-NS mergers, we note that their formation mechanisms and timescales are so similar to NS-NS mergers that we can consider them part of the same process, though we note from \cite{Pannarale2014} that for $M_\text{BH} \gtrsim 14 M_\odot$, the vast majority of NS companions will be swallowed without tidal disruption, and hence without r-process ejecta.

	\subsection{Simple Argument}\label{S:Simple}
	
		Before deriving our analytical chemical evolution models, it is instructive to first discuss the qualitative appearance of Collapsar-dominated models in both the [Eu/Fe]-[Fe/H] and [Eu/Mg]-[Fe/H] planes.
		
		In agreement with previous works (i.e. \siegel{}), we parametrise r-process yields from collapsars (see eq. \ref{E:FullYields}) with a constant synthesis rate for metallicity $Z < \Zc$. This assumption guarantees that the dominant collapsar pathway produces a plateau in both the [Eu/Fe] and [Eu/Mg] planes. These approximations are justified by comparison with the data in Figure \ref{F:SAGAData}, which shows a plateau in both planes at early times. 
		
		Since the SNIa channel opens with a time delay and produces mostly iron, the plateau in the [Eu/Fe] plane will be interrupted, producing the sharp dip in the [Eu/Fe] abundance (the knee) as the iron production increases. The [Eu/Mg] abundance, however, will exhibit only minor changes (due to metallicity variations or the subdominant Mg yield from SNIa), matching the observations of Fig. \ref{F:SAGAData}.
		
		When the background metallicity reaches $Z \sim \Zc$, the Eu synthesis rate will drop as collapsars cease to be formed. Iron and Magnesium production, however, undergo no such transition. This will manifest a `knee' in both the [Eu/Fe] and [Mg/Fe]  planes, with both abundance tracks dropping significantly from their previous values. However, observational evidence for such a feature is lacking. Fig. \ref{F:SAGAData} does not show a second knee in the iron plane, and nor is there evidence of a significant drop in [Eu/Mg] abundance -- in fact, [Eu/Mg] remains almost flat across the entire space.
		
		This simple argument already appears to put severe constraints on the dominance of collapsars in the production of r-process elements. Note, however, that whilst these naive constraints hold in general, they can be modified, e.g. by radial migration \citep{Tsujimoto2019}, the variations in the SFH, and the cooling of hot gas.

	\subsection{Analytical Model}\label{S:Model}
		
		The purpose of this paper is to explore the full possible parameter space of collapsar contributions to examine which regions are ruled out, not to point to some plausible solutions. This necessitates a rapid, streamlined model which can evaluate the chemical histories for billions of combinations of possible parameters. 
		
		To this end, we developed a Simple Analytical Chemical Evolution Model  (\modelname{}), which makes a number of simplifying assumptions to allow a fully analytical solution. We use this model to track the enrichment of three elements representative of three key groups: the alpha elements (Mg), the iron peak (Fe) and r-process (Eu). 
		
		\modelname{} features many aspects common to standard GCE models, and we limit the following discussion to the points distinguishing it from other models. More details are found in Appendix \ref{A:Model}.
		
		\modelname{} is a single-zone model but features two gas phases: a cold phase capable of forming stars and a hot phase into which the majority of newly synthesised material is ejected. This follows the work of \cite{Schonrich2019} which showed that such thermal splitting had a drastic impact on the chemical histories inferred for r-process material, and allows us to incorporate effects such as the diffuse return of hot gas to the star-forming disc: the `galactic fountain' of \cite{Shapiro1976}.
	
		\subsubsection{Modelling r-Process Synthesis}\label{S:CollYields}
		
			The rate at which collapsars synthesise r-process material, $y_\text{r}(t)$, is dependent on the properties of the progenitor star (mass, $M$, metallicity, $Z$, and ZAMS rotation speed, $v$), the r-process yield $Y_r(M,Z,v)$ of each event from such a progenitor and the rate at which these progenitors undergo a collapsar event, $R_\text{coll} = \d N_\text{coll} / \d t \d M \d Z$, such that the exact synthesis rate is:
			
			\begin{equation}
			y_\text{r}(t) = \int_0^\infty\d M \int_0^1 \d Z \int_0^\infty \d v ~Y_r(M,Z,v) R_\text{coll}(M,Z,v,t) \label{E:ExactYield}
			\end{equation}
			
			In \S\ref{S:CollYieldBasics} we identified several problems barring us from formulating $Y_r$ and $R_\text{coll}$, so we must instead opt for simple approximations to \eref{E:ExactYield}. Consistent with the approach of \siegel{} we assume that the yields from individual events are a constant, and that collapsars are a subset of CCSN events such that a fraction $\xi_\text{coll} a(Z,\Zc)$ of all stars undergoing CCSN meet the criteria to become collapsars. The constant $\xi_\text{coll}$ accounts for the mass and rotational constraints, whilst $a(Z,\Zc)$ is the metallicity suppression function which obeys $a(Z,\Zc) = 0$ for $Z > \Zc$. Therefore:
			
			\begin{equation}
			y_\text{r}(t) \approx \xi_\text{coll} \bar{Y}_\text{coll} \times R_\text{CCSN}(t) \times a(Z(t), \Zc)\label{E:FullYields}
			\end{equation}
			
			Here $\bar{Y}_\text{coll}$ is the constant characteristic yield of a single collapsar event, $R_\text{CCSN}(t)$ is the rate of CCSN events at the time $t$ and $Z(t)$ is the metallicity of the star-forming gas at a time $t$. Note that although the ejecta-mass from a single collapsar is assumed to be a constant, due to the metallicity dependence of the collapsar rate, the total synthesis rate is strongly metallicity dependent.

			The formulation of reasonable functions $R_j(t)$ for different processes and $a(Z)$ is discussed in detail in Appendix \ref{A:Model}, whilst the formulation of $Z(t)$ is found in \S\ref{S:MetalDecouple}.

		\subsubsection{Metallicity Decoupling}\label{S:MetalDecouple}
	
			\begin{figure}
			    \centering
			    \fittedImage{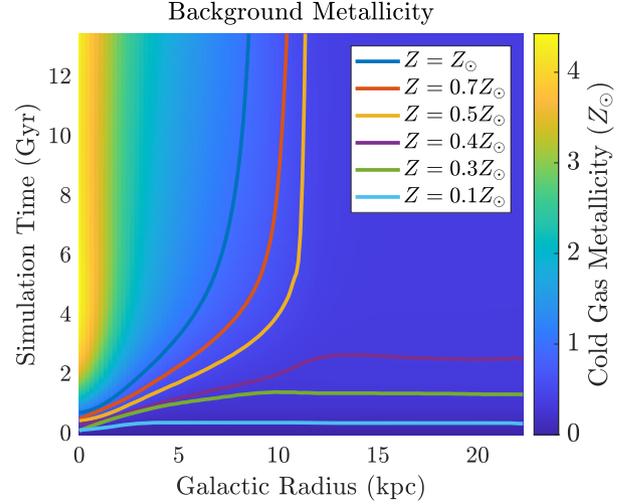}
			    \caption{A plot showing the background metallicity evolution from the best-fit \codename{} simulation. Superposed onto this are contours of constant metallicity, which are used to derive the critical cutoff time in \modelname{}}
			    \label{F:Background}
			\end{figure}
			
			\modelname{} tracks the abundance of 3 chemical species, but not the overall metallicity. However, it is known that the yields of single elements can show strong and individual metallicity dependencies \citep{Maeder1992,Chieffi2004}, making $Z(t)$ a functional of its past evolution which is impossible to evaluate analytically. For simplicity, we therefore fix $Z = \Z(t)$, where $\Z(t)$ is an externally imposed, monotonic function of time. As seen in Appendix \ref{A:Model}, the only explicitly metallicity dependent part of our model is the suppression rate of collapsar yields, which `turns off' at $Z = \Zc$. Utilising the monotonicity of $\Z(t)$, we therefore instead suppress collapsar contributions when $t = \tc$, such that:
			
			\begin{equation}
				\Z(\tc) = \Zc
			\end{equation}

			We choose $\Z(t)$ to be the function recovered from an instantiation the full \codename{} simulation (\S\ref{S:Ramices}) which, as per \sbRef, replicates many aspects of the historic enrichment profile. The evolution of $\Z(r,t)$ from the best-fit \codename{} model is shown in Fig. \ref{F:Background}. Since our data is mostly from the solar neighbourhood, we evaluate this function at the solar radius, which is well represented for $0 < t < 14$Gyr by the polynomial:
	
			\begin{equation}
				\Z(t \text{ Gyr}) = Z_\odot \left( 0.000591 t^3 -0.0205 t^2  + 0.255 t - 0.00459 \right) \label{E:ZFit}
			\end{equation}
			
		\subsubsection{Deviating from the Metallicity Enrichment History}\label{S:EnrichmentHistory}
	
			Varying the properties of our model galaxies will cause the subsequent chemical history to diverge from that used to generate \eref{E:ZFit}. This is potentially highly problematic for us, as we are now imposing $\tc$ as a model parameter, instead of the physically meaningful value $\Zc$. 
			
			It is possible, therefore, that a model successfully replicates the r-process enrichment history of the Milky Way, and has a true cutoff metallicity $Z_t$ such that $Z_t < 0.3 Z_\odot$. However, because we have not updated $\Z$ for this model, we find that $\Z(\tc) > 0.3 Z_\odot$, and we would reject this model as `unphysical'.   
			
			However, by reference to i.e. \cite{Casagrande2011,Schonrich2017,Haywood2008}, we have constraints on what a physically meaningful $Z(t)$ can be. Since \codename{} (and hence Eq. \ref{E:ZFit}) is calibrated specifically to reproduce this information, the observational evidence constrains how much $\Z(\tc)$ and $Z_t$ can differ. The `falsely rejected' models above are explicitly those which lie in tension with this data. 
			
			Any model which has $\tc \gtrsim \tc(\Zc)$ can therefore be rejected for either: 
			
			\begin{itemize}
			\item Having $Z_t > 0.3 Z_\odot$, and therefore failing the test of \S\ref{S:Cutoff}
			\item Requiring a metallicity enrichment history which contradicts observed evidence
			\end{itemize}
			
			The converse is also true: we may generate models with $Z_c(\tc) < 0.3 Z_\odot < Z_t$. These models would pass the test of \S\ref{S:Cutoff}, but are unphysical. However, due to our focus on negative inference, accepting unphysical models limits our conclusions to upper bounds on the contribution of collapsars. Of far greater concern to us is rejecting physical models, which we must ensure we do not do.
	
		\subsubsection{Yield Calibration}\label{S:Calibrate}
			
			Due to the parameterisation of the stellar yields, each pathway has an undetermined constant in the form of an effective yield $\bar{Y}_{Xj}$, usually a result of the IMF-weighting of the true yield, as well as effects such as galactic-ejection fractions and remnant-lockup.
			
			A true GCE model (e.g., \citealt{Portinari2000}) would try to derive these prefactors from first principles - for the purposes of \modelname{}, we instead fix these values by requiring that the curves of interest to us (those for Fe, Eu and Mg) replicate some chosen calibration values. The chosen calibration points, and their values in the nominal model, are shown in Table \ref{T:Calibration}. 
			
			We note again that this strategy allows additional degrees of freedom into our model, but that since we are investigating which regions of parameter space are excluded, this in fact strengthens any conclusions we might draw: additional constraints on $\bar{Y}_{Xj}$ would serve to reduce the viable parameter space, not expand it. 
			\newcommand\cLine[4]
			{
				#1	(at $t=$#2)	&	$#3$	&	#4	\\
			}
			\def\today{$t_\text{today}$}
			\begin{table}
				\caption{The calibration points used in the \modelname{} to fix the yields of CCSN, SNIa, Collapsar and NSM events. The final constrant, $\chi$, has no nominal value as it the parameter we wish to investigate: the contribution of collapsars to the total r-process budget. }\label{T:Calibration}
				
				\small
				\begin{tabular}{c c c c}
				\bf Observable &	\bf Symbol &	\bf Nominal Value
				\\ \hline
				\cLine{ $[$Fe/H$]$}{\today}{\mathcal{F}}{0.1 dex}
				\cLine{ $[$Mg/Fe$]$}{\today}{\mathcal{M}_\infty}{0 dex}
				\cLine{ $[$Mg/Fe$]$}{$\tau_\text{SNIa}$}{\mathcal{M}_0}{0.35 dex}
				\cLine{ $[$Eu/Fe$]$}{\today}{\mathcal{E}}{0 dex}
				\cLine{s Process fraction}{\today}{\sigma}{0.02}
				\cLine{Collapsar fraction}{\today}{\chi}{-}
				\hline
				\end{tabular}
			\end{table}

		\subsubsection{Action of Delays \& Thermal Phasing}\label{S:Phasing}
		
		\begin{figure*}
			    \centering
			    \fittedImageLarge{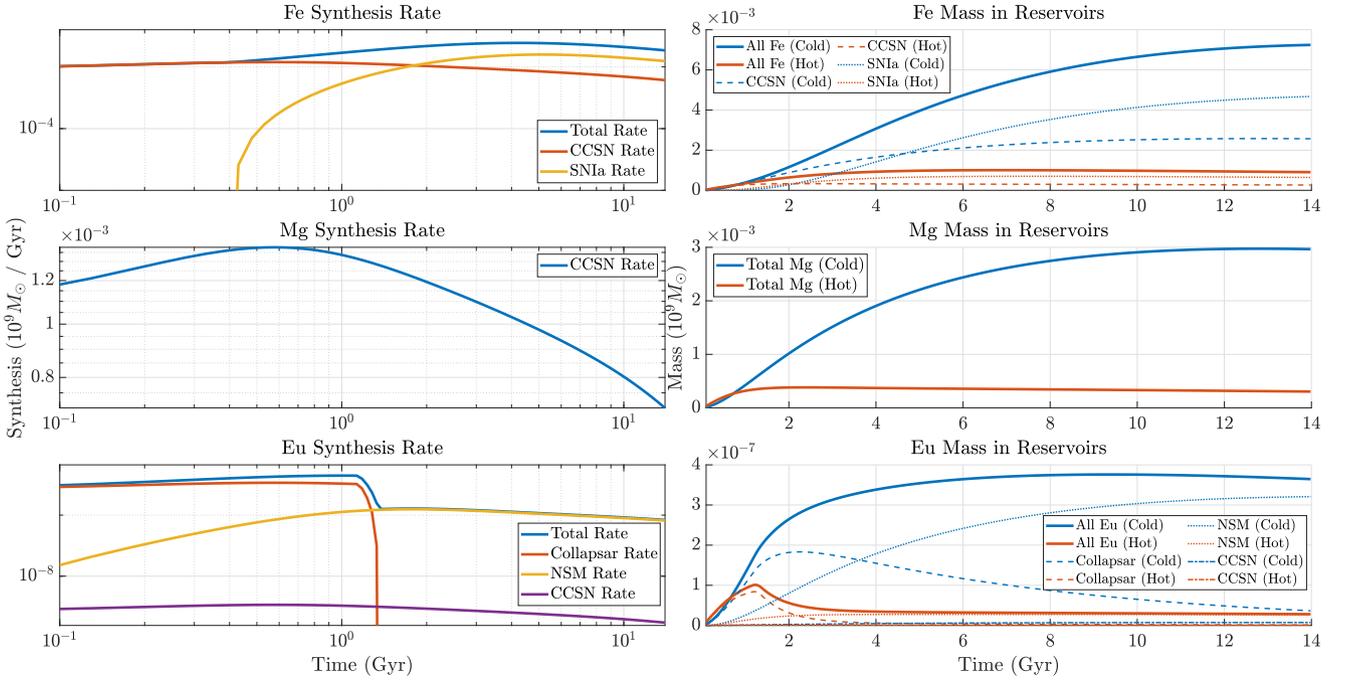}
			    \caption{ {\it Left: } The rate of synthesis of the three species within \modelname{} and {\it right:} the resulting mass of each element stored in the hot and cold gas phases for a calibrated \modelname{} instantiation chosen to generate r-process contributions $(\chi,\sigma,\kappa) = (0.1,0.02,0.88)$ at simulation end, and with a collapsar cutoff time $\tc = 1.35$Gyr, corresponding to $\Zc \approx 0.3Z_\odot$. The values of $(\chi,\sigma,\kappa)$ were selected to amplify the signal of the differing behaviours, rather than generate a viable chemical history. This model would be considered `unsuccessful' by \S\ref{S:Success} - though we note that a qualitatively identical (but harder to interpret) plot for $(\chi,\sigma,\kappa)=(0.02,0.02,0.96)$ would be considered successful.}
			    \label{F:Synthesis}
		\end{figure*}	

		Central to understanding our chemical evolution models are the differing Delay-Time Distributions (DTD) of yields, and delays of injection to and freeze-out from the hot gas phase.  

		For the DTDs, CCSN and Collapsars occur almost immediately as the lifetime of high-mass stars is negligible compared to the timescale of chemical evolution (~10Gyr). SNIa and NSM, however, rely on the death of a previous population of stars for their formation mechanism, giving a minimum timescale of about 200Myr for SNIa and 10Myr for NSM. With reference to \S\ref{S:Cutoff}, we also have that the collapsar channel closes at the time $\tc$, as the metallicity at this time is too large to allow collapsar formation.
		
		The left hand panels of Fig. \ref{F:Synthesis} shows the (calibrated) rates at which the three tracked metals are synthesised within a given \modelname{} initialisation, demonstrating the impact that these differing timescales and cutoffs have on the associated yield functions. 
	
		In addition to the absolute rate of synthesis, \cite{Schonrich2019} showed that the rate at which polluted gas becomes available for star formation also has a large impact on GCE models of r-process synthesis, because we expect diffrent timescales for CCSN, NSM and SNIa gas-availability. For example: the majority of CCSN occur close to regions of high star formation (and hence close to the feedback `chimneys' of \citealt{Norman1989}), so we expect a large contribution of the gas to be stored into the hot reservoir, or temporarily ejected from the galaxy, only to `fountain' back in, as per \citep{Shapiro1976}. NSM, however, have a significant time delay, and also experience natal kicks, so can be expected to be far removed both spatially and temporally from feedback chimneys, and their synthesis of high-mass, neutron rich material would lead to strong line cooling of the ejecta, making the material available for star formation much more rapidly. Microphysics such as dust-formation may also have an impact on the availability of gas for star formation.
		
		To model these effects, \modelname{} places a portion $f_j$ of the material synthesised by process $j$ into the hot gas reservoir, with the remainder going into the hot-gas reservoir \footnote{\modelname{} does not include a total-ejection fraction as this takes the form of a multiplicative prefactor $y_\text{non-eject} = (1 - f_\text{eject}) y_\text{X}$, as $y_\text{X}$ is calibrated (\S\ref{S:Calibrate}), we can omit the factor and calibrate the value of $f_\text{eject}$ simultaneously with the absolute yield.}. We then allow the each process material to cool at a slightly different rate such that  $\dot{M}_\text{cool} =\lambda_j M_\text{hot}$. Following \cite{Schonrich2019}, we adopt $(f_\text{CCSN}, f_\text{Coll}, f_\text{NSM}, f_\text{SNIa}) = (0.75,0.75,0.4,0.99)$ and $\lambda_j= \lambda_\text{CCSN} = 1\text{Gyr}$ as our nominal model, though we note that their fixing of the non-NSM values was somewhat arbitrary. The right hand panels of Fig. \ref{F:Synthesis} show how the synthesised yields are injected, cooled and consumed by star formation over the course of galactic history. 
		
		In this class of models the `dominant' source of Eu changes drastically with time: even though the final contribution of collapsars to the Eu budget is $\sim 10 per cent$, at early times it accounted for more then 90 per cent of the synthesised Eu, of which more than 50 per cent was found within the hot gas phase, and the collapsar-dominant early time contribution was maintained even as $\chi < 0.01$. Even if collapsars are negligibly responsible for r-process synthesis at late times, Fig. \ref{F:Synthesis} shows that they may have been important contributors at early times. This model class would be disfavoured due to the \sneden{} fingerprint indicating a monolithic r-process source, however, if we suggest that the high-[Eu/Fe] stars sampled to form the `fingerprint' may not be representative of the bulk of stars at this early time (which did source their Eu from collapsars), this tension is alleviated.

		\subsubsection{Resolving Contribution Ambiguity}
	
			$\chi$, the collapsar contribution fraction, has two distinct definitions:
			\begin{enumerate}
				\item The mass ratio of collapsar-sourced europium to the total europium within the cold, star forming gas at a time $t$
				\item The total mass ratio of all collapsar-sourced material ever produced to the total amount of europium ever synthesised (i.e. including that which has been subsequently locked up in stars, stored in the hot gas reservoir or folded into black holes)
			\end{enumerate}
			
			These two definitions can diverge significantly: material produced early on is either lost to the IGM or locked up in stars/stellar remnants, so the definition (i) sets weights much more towards more recent enrichment. With that significant difference, we note that the qualitative picture and our inferred conclusions remains unaltered between the two choices and we will use option (i), defining $\chi$ as the collapsar contribution to the current europium in the cold gas phase. The s-process contribution, $\sigma$ and NSM contribution, $\kappa$ are defined similarly.

		\subsubsection{\modelname{}: A Recipe}\label{A:Recipe}
	
			With reference to the derivations in Appendix \ref{A:Model}, we set out a procedure to derive a chemical history as a function of the parameters of a given \modelname{} model (listed in table \ref{T:Parameters}):
			
			\begin{enumerate}
				\item Solve differential equations \eref{E:McDot}-\eqref{E:MtDot} to generate a SFR:
				\begin{equation}
				\rho_\text{SFR}(t) = \rho(t | M_0, M_1, M_2, \beta_1, \beta_2, \nu_\text{sfr}, \mu, \delta)
				\end{equation}
				\item Use the SFR to generate the (uncalibrated) event rates for the four processes, using $\mathcal{I}$ from \eref{E:DelayYield}:
				\begin{align}
					R_\text{ccsn}(t, \rho_\text{sfr}) &= \rho_\text{sfr}(t)
					\\
					R_\text{colls}(t,\rho_\text{sfr}, \tc, \Delta) & = \rho_\text{sfr}(t) a(t,\tc,\Delta)
					\\
					R_\text{k}(t,\rho_\text{sfr}, \nu_\text{k}, \tau_\text{k}) & = \mathcal{I}\left[\rho_\text{sfr}, \nu_\text{k}, \tau_\text{k}\right]~~k\in\{\text{snIa},\text{nsm}\}
				\end{align}
				\item Solve \eref{E:ColdGasOrigin}-\eqref{E:HotGasOrigin} to find the (uncalibrated) cold-gas mass in the disc produced by each process as a function of the relevant yield, thermal and lockup parameters.
				\begin{align}
					M_k & = M\left[t, R_k, f_{h,k}, \lambda_k, \delta, \nu_\text{SFR}, F_\text{mod} \right]
					\\
					&~~~k\in\{\text{ccsn, colls, nsm, snIa}\}
				\end{align}
				\item Construct models for the mass of each element within the cold gas as a function of unknown prefactors:
				\begin{align}
					M_\text{Fe} & = \alpha M_\text{ccsn} + \beta M_\text{snIa}
					\\
					M_\text{Mg} & = \gamma M_\text{ccsn}
					\\
					M_\text{Eu} & = \delta M_\text{ccsn} + \epsilon M_\text{coll} + \zeta M_\text{nsm}
				\end{align}
				\item Invoke a function $\mathcal{C}$ which calibrates the unknown prefactors against the observed data and model inputs of Table \ref{T:Calibration}:
				\begin{align}
					\tilde{M}_\text{H}(t) & = X M_c = X \rho_\text{SFR}/\nu_\text{SFR}
					\\
					\tilde{M}_\text{Fe}(t) &= \mathcal{C}(M_\text{Fe}, M_\text{Mg}, \tilde{M}_\text{H}, \mathcal{F}, \mathcal{M}_\infty)
					\\
					\tilde{M}_\text{Mg}(t) &= \mathcal{C}(M_\text{Mg}, \tilde{M}_\text{Fe}(t),\mathcal{M}_0)
					\\
					\tilde{M}_\text{Eu}(t) &= \mathcal{C}(M_\text{Eu}, \tilde{M}_\text{Fe}(t),\mathcal{E},\sigma,\chi)
				\end{align}
			\end{enumerate}
			This method generates 4 analytical functions which can be used to plot Tinsley diagrams of the generated chemical history of the galaxy, and forms the core of \modelname{}.

	\subsection{Full Simulation} \label{S:Ramices}
	
		\modelname{} is designed to run quickly with minimal resources, allowing for maximal parameter searches. The penalty for this, however, was a number of potentially unpalatable approximations. In order to ensure that \modelname{} is not leading us astray, we also make use of a full multi-zone, multi-phase GCE model which captures much more physics - at the cost of orders of magnitude more computation time. 
		
		We use a modified version of of the Radial Migration with Chemical Evolution Simulation (\codename{}) code developed by \cite{Schonrich2009} (\sbRef), with the updated parameters and inside-out disc growth developed in \cite{Schonrich2017} and the dual-phase NSM r-process injections of \cite{Schonrich2019}. 
		
		A brief description of the code and its functionality can be found in Appendix \ref{A:Simulation}, though the interested reader may find the above papers more complete. 
		
		For the present work, we modified the base model: we have expanded and updated the \codename{} chemical yield grid to account for a collapsar contribution (modelled simply as a subset of CCSN events) and a low level s-process contribution to the europium synthesis. In addition, we can now track elements by source - allowing us to distinguish between NSM-origin metals and collapsar-origin.

\section{Search For Models} \label{S:GridSearch}

	\newcommand\mRow[9]
			{
				~ & $#1$	&	\parbox[t]{6.4 cm}{#2}	& #9 &	#3	&	#4	&	#5	&	#6 & #7 &#8\\
			}
	\newcommand\insertBigTable
	{
	\def\massU{$10^{9} M_\odot$}
	\def\freqU{Gyr$^{-1}$}
	\begin{table*}
	
		\caption{A list of the named parameters within \modelname{}, their physical definitions and interpretations, and the bounds placed on them in the three primary constraint sets. Note that the mixed set, \texttt{Set M}, uses the constraints from the Viable set with the exception of the `SFR Parameters', which it draws from the unconstrained set. For an algorithmic `recipe' for how to incorporate these parameters into \modelname{}, see \S\ref{A:Recipe}} \label{T:Parameters}
		\begin{center}
			\small
		
			\begin{tabular}{|c c p {5.8cm} c @{\vline}c c @{\vline}c c @{\vline}c c }
			~ & \bf Quantity&	\bf Description &	\bf Units & \multicolumn{2}{c}{\bf Unconstrained}	&	\multicolumn{2}{c}{\bf Weak Constraints}		&	\multicolumn{2}{c}{\bf Viable Constraints}	
				\\
				~& ~	& ~ &~& \scriptsize Lower  & \scriptsize Upper ~&\scriptsize~ Lower &\scriptsize  Upper~~&~\scriptsize Lower &\scriptsize  Upper
				\\ \hline
				\mRow{M_0}{Initial Cold Gas Mass present in galaxy}{0.001}{200}{0.01}{10}{0.1}{1}{\massU}
				\mRow{M_1}{Primary Infall Mass (thick disc)}{$10^{-6}$}{20}{1}{10}{1}{10}{\massU}
				\mRow{M_2}{Secondary Infall Mass (thin disc)}{$10^{-6}$}{200}{10}{70}{20}{70}{\massU}
				\mRow{\beta_1}{Primary Infall Frequency}{0.1}{100}{0.1}{10}{0.333}{10}{\freqU}
				\mRow{\beta_2}{Secondary Infall Frequency}{0.001}{1}{0.02}{0.1}{0.033}{0.1}{\freqU}
				\mRow{\nu_\text{SFR}}{Star Formation Rate frequency}{0.01}{8}{0.05}{5}{0.05}{5}{\freqU}
				\mRow{\mu}{Stellar Death Frequency (rate = $\mu M_*$)}{$10^{-4}$}{1}{0.001}{0.5}{0.001}{0.1}{\freqU}
				\mRow{\delta}{Outflow/heated mass per stellar mass formed}{$10^{-5}$}{5}{0.01}{2}{0.01}{1.5}{-}
				\multirow{-9}{*}{\rotatebox[origin=c]{90}{\it SFR Parameters}}\mRow{M_c/M_*}{Final ratio of cold mass to stellar mass}{$10^{-5}$}{1}{0.05}{0.25}{0.05}{0.15}{-}
				\hline
				\mRow{f_\text{CCSN}}{Fraction of CCSN mass put into hot phase}{0.001}{0.999}{0.6}{0.999}{0.7}{0.999}{-}
				\mRow{f_\text{NSM}}{Fraction of NSM mass put into hot phase}{0.001}{0.999}{0.3}{0.999}{0.3}{0.999}{-}
				\mRow{f_\text{SNIa}}{Fraction of SNIa mass put into hot phase}{0.001}{0.999}{0.6}{0.999}{0.7}{0.999}{-}
				\mRow{f_\text{Coll}}{Fraction of Collapsar mass put into hot phase}{0.001}{0.999}{0.6}{0.999}{0.7}{0.999}{-}
				\mRow{\lambda_\text{CCSN}}{Cooling frequency for ejecta from CCSN}{0.04}{10}{0.4}{2.5}{0.5}{1.5}{\freqU}
				\multirow{-6}{*}{\rotatebox[origin=c]{90}{\it Thermal Parameters}}\mRow{\ell_j}{3 independent parameters. $\lambda_j= \lambda_\text{CCSN} ( 1 + \ell_j)$ \\for $j\in\{\text{CCSN, SNIa, NSM}\}$}{-0.999}{0.999}{-0.2}{0.2}{-0.1}{0.1}{-}
				\hline
				\mRow{\tau_\text{SNIa}}{SNIa delay time}{0.001}{2}{0.05}{1}{0.1}{0.6}{Gyr}
				\mRow{\tau_\text{NSM}}{NSM delay time}{$10^{-5}$}{2}{$10^{-4}$}{0.6}{$10^{-3}$}{0.1}{Gyr}
				\mRow{\nu_\text{SNIa}}{SNIa DTD Decay frequency}{0.05}{50}{0.05}{25}{0.1}{15}{\freqU}
				\mRow{\nu_\text{NSM}}{NSM DTD Decay frequency}{0.05}{50}{0.05}{25}{0.05}{25}{\freqU}
				\multirow{-10}{*}{\rotatebox[origin=c]{90}{\parbox[t]{3 cm}{\it Temporal \\ Parameters}}}\mRow{\Delta}{Collapsar turnoff width}{0.1}{15}{0.1}{15}{0.1}{15}{Gyr}	
				\hline
				\mRow{F_\text{mod}}{Lockup-SFR modification factor}{0.3}{1}{0.3}{1}{0.3}{1}{-}
				\mRow{X}{Hydrogen Fraction at simulation end}{0.5}{0.9}{0.65}{0.75}{0.68}{0.72}{-}
				\mRow{\mathcal{F}}{Final [Fe/H] value}{-0.5}{1}{0}{0.5}{0.05}{0.3}{dex}
				\mRow{\mathcal{M}_0}{Thick disc value for [Mg/Fe]}{0.2}{0.6}{0.3}{0.5}{0.3}{0.4}{dex}
				\mRow{\mathcal{M}_\infty}{Final value for [Mg/Fe]}{-0.3}{0.2}{-0.1}{0.1}{-0.1}{0}{dex}
				\mRow{\mathcal{E}}{Final value for [Eu/Fe]}{-0.3}{0.2}{-0.1}{0.1}{-0.1}{0.05}{dex}
				\mRow{\sigma}{Fraction of europium produced through the s process at simulation end}{$10^{-8}$}{0.2}{$10^{-8}$}{0.1}{$10^{-8}$}{0.05}{-}
				\multirow{-11}{*}{\rotatebox[origin=c]{90}{\parbox[t]{3 cm}{\it Chemical \\ Parameters}}}
				\mRow{\kappa}{Fraction of europium produced through NSM \\$\kappa \equiv 1 - \sigma - \chi$}{-}{-}{-}{-}{-}{-}{-}
				\hline
				\mRow{\tc}{Time of final collapsar event}{0}{16}{0}{16}{0}{16}{Gyr}
				\mRow{\chi}{Fraction of Europium produced in collapsars at simulation end}{0}{1}{0}{1}{0}{1}{-}
						\hline
			\end{tabular}
		\end{center}
		
	\end{table*}
	
	}
	
	\subsection{Varying $\tc$ in Analytical Models} \label{S:DumbSearch}

		As an initial experiment (and to confirm the intuition developed in \S \ref{S:Simple}), we observe the effects of modifying the collapsar cutoff time, $\tc$ on \modelname{} chemical histories. 
		
		The investigated model was calibrated against the iron and magnesium distributions - the only constraints placed on the Europium abundance is that, at simulation end, [Eu/Fe] = 0 and $(\chi,\sigma) = (0.98,0.02)$, i.e. be 98 per cent collapsar in origin, with no NSM contribution.
		
		Using the $Z_\text{cut}-\tc$ relationship of \S\ref{S:MetalDecouple}, we generate 6 chemical histories corresponding to cutoff metallicities of 0.1, 0.3, 0.5, 0.7, 1 and 20 $Z_\odot$, with the final cutoff being deliberatly large so as to take place at an infinite time in the future (we refer to this model as `$\Zc \gg Z_\odot$'). 
		
\begin{figure}
		 \centering
			\fittedImage{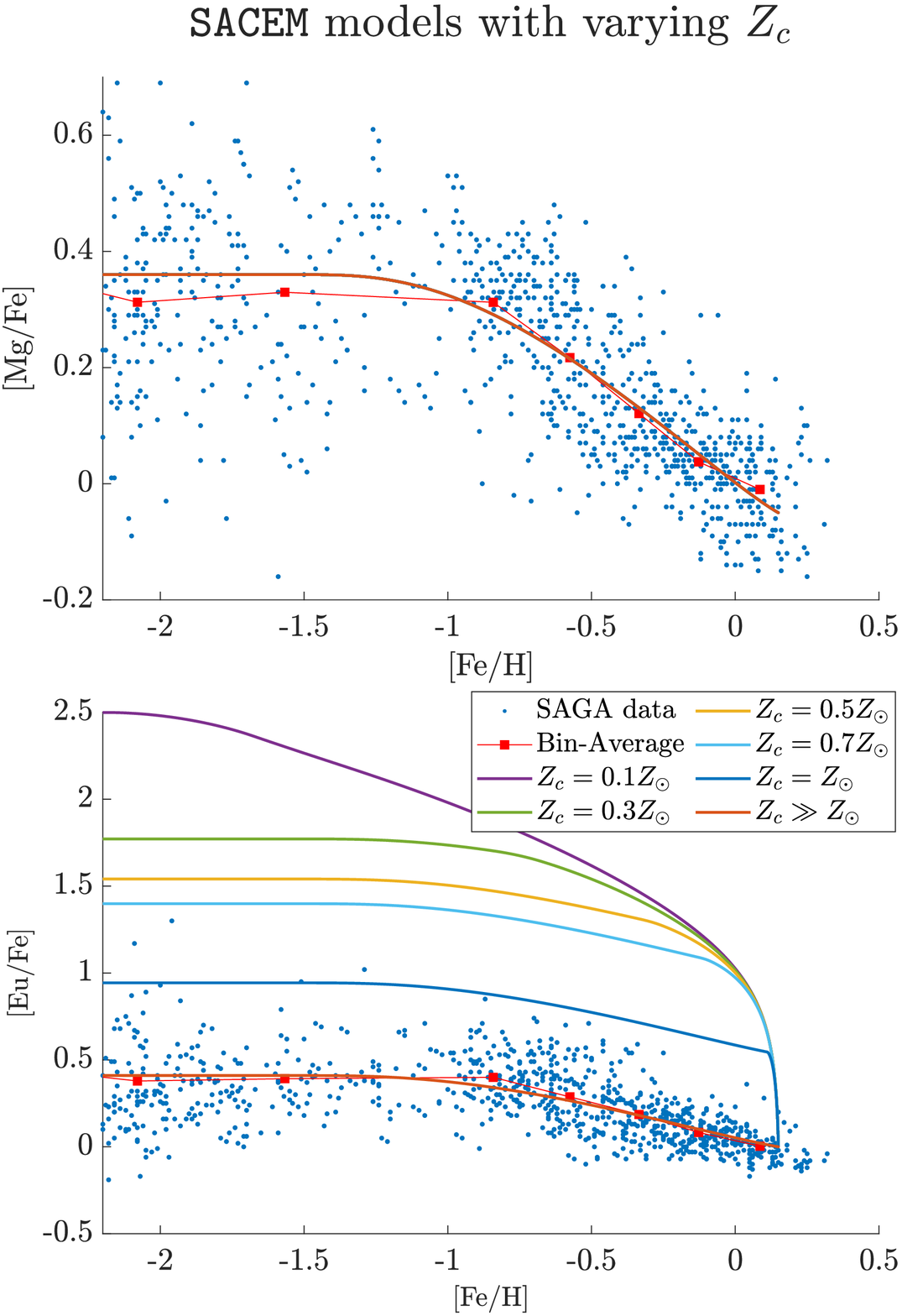}
			 \caption{A series of chemical histories generated by \modelname{} for a collapsar-dominated galaxy, for various values of the collapsar cutoff time $T = \tc$. The time/metallicity relationship is calculated from Fig. \ref{F:Background}.}
		    \label{F:MultiSACEM}
		\end{figure}

		The resultant chemical histories, shown by the solid lines in Fig.\ref{F:MultiSACEM}, differ only in the normalisation of their Eu channels, so the behaviour in [Mg/Fe]-[Fe/H] (top panel) is unaffected. However, in the lower panel we see the impact in the [Eu/Fe]-[Fe/H] plane: since the models are tethered to reach [Eu/Fe] $\sim$ 0 at the end of the simulation, \modelname{} compensates for the truncated Europium production by increasing the collapsar effective yield $\bar{Y}_\text{Coll, Eu}$, resulting in unreasonably large [Eu/Fe] values at early times: these models cannot be accurate depictions of our galaxy.

		\begin{figure}
		 \centering
			\fittedImage{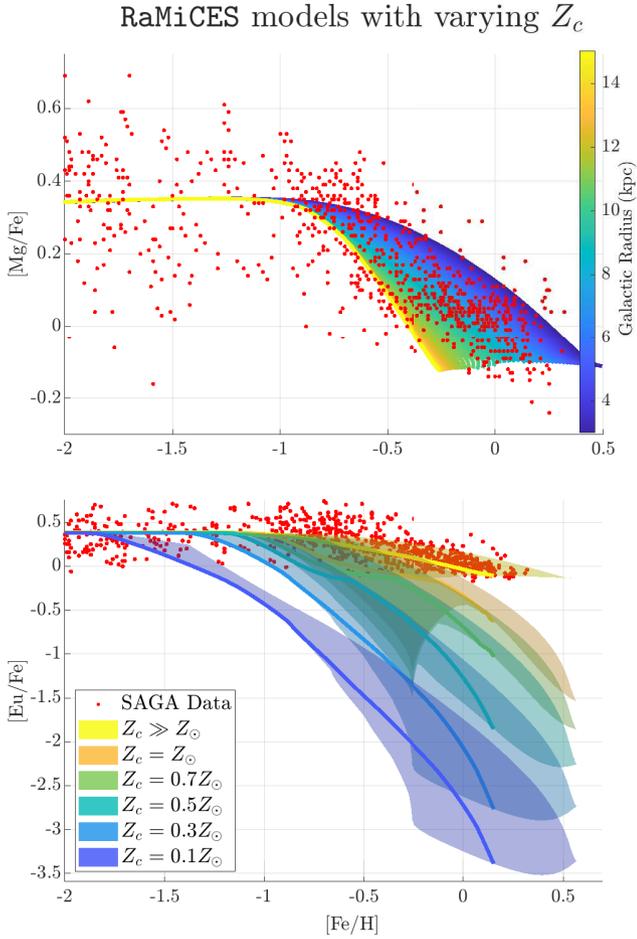}
			 \caption{As with Fig. \ref{F:MultiSACEM}, but generated using the \codename{} simulation. In the top panel, the color is used to denote the galactic radius. In the bottom panel, the shaded region shows the same radial distribution as in the top panel with the thick line tracking the chemical history at the solar radius.}
		    \label{F:RamicesBacktrack}
		\end{figure}

	\subsection{\codename{} \& Radial Structure}
	
		This is reinforced by Fig \ref{F:RamicesBacktrack}, in which we also run the same set of constraints on the \codename{} simulation, verifying that these initial results hold up in the full multi-zone model. Unlike \modelname{}, the models used to produce Fig. \ref{F:RamicesBacktrack} are calibrated to match the early-time paths, and so we see these curves plunge below [Eu/Fe] = 0, rather than their early-time abundances shooting up. 
		
		We see that the \codename{} models suffer from a greater differential between their early time abundances and their final values than \modelname{} did. This indicates that the approximations in \modelname{} make it more lenient with regards to gas depletion timescales, or late-time Eu abundances sourced from the hot gas phase. As a result, \modelname{} will provide less stringent constraints for the late Eu production than the full simulation. This works in our favour, as it means any regions of parameter space that we are able to exclude is likely to be a lower bound on the size of the excluded region. 
		
		\begin{figure}
		 \centering
			\fittedImage{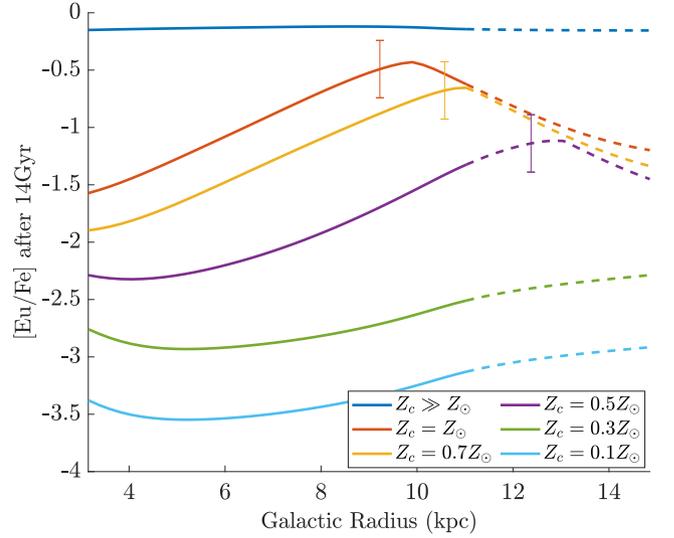}
			 \caption{The final result of the simulations of Fig. \ref{F:RamicesBacktrack} at $t = t_\text{end}$ explicitly projected into radial coordinates. The lines are dashed from the point at which the 50 per cent of the total metallicity budget was accounted for by the IGM. Where present within the galactic disc, the vertical lines denote the physical position of the `critical radius': the point where $\Z(r,t_\text{end})= \Zc$ 
				 }\label{F:RadialPlot}
		\end{figure}
		As a multi-zoned model, \codename{} allows us to investigate the guiding-centre radius ($R_g$) behaviour of the abundance distributions, shown in Fig. \ref{F:RadialPlot}. The rise in [Eu/Fe] seen from $R_g = 3$ to $R_g \sim 10$kpc would be what we expect from the impact of the galactic metallicity gradient: the inner galaxy hits $\Z = \Zc$ first, and so r-process abundances begin to plummet first at lower $R$. The inner radii have been Europium-deprived for longer than mid-disc radii, and so have a lower [Eu/Fe] ratio - hence the positive [Eu/Fe] gradient as you move out through the disc. 
		
		For $Z = 0.5,~0.7$ and $1Z_\odot$, this pattern in [Eu/Fe] is interrupted by an `arch' at $R_g > 10$kpc - an effect induced by the inflow/accretion prescriptions of \codename{}. Currently, \codename{} tethers the composition of infalling material to the composition of the gas found at a certain galactocentric radius. The `arch' indicates the point at which accreted material becomes the dominant driver of the cold phase metallicity - and since the abundances of the inflowing material cannot yet be reliably determined, this indicates the point at which \codename{} cannot robustly predict the radial structure.  
		
		The behaviour inwards of the solar radius, however, is robust against the IGM prescription chosen and is a necessary consequence of the metallicity gradient of the galaxy. Though a lengthier discussion of the radial behaviour of [Eu/X] is beyond the scope of this paper, we note that the radial distribution of europium, as inferred from cepheids in \cite{Luck2011} is inconsistent with a such a drastic increase in [Eu/Fe] with galactic radius. This further emphasises the conclusions drawn in \S \ref{S:TimeData}: there is little evidence for a large change in galactic chemistry in the recent past - any changes to the galactic chemistry must have been far enough in the past for the radial mixing of the galaxy to smooth out the emergent patterns.  
			\begin{figure}
		 \centering
			\fittedImage{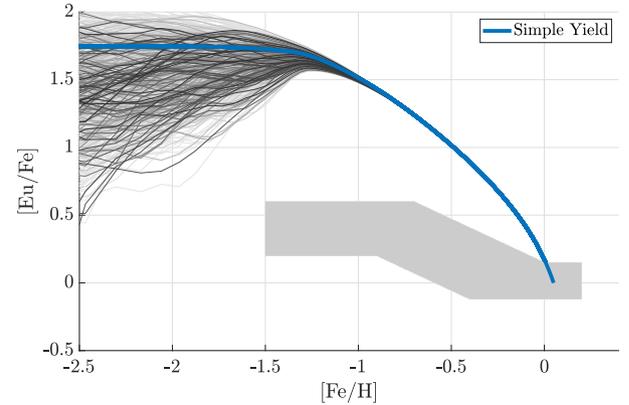}
			 \caption{10,000 SACEM instantiations in which the collapsar yield function is allowed to take on arbitrary forms are shown, relative the the grey `allowed region'. All models have $\tc = 1.4$Gyr, $\chi = 1$ and otherwise use the best fit parameters of Fig. \ref{F:MultiSACEM}. We see that within a fraction of a dex of [Fe/H] after collapsar suppression, all models converge on a single path through [Eu/Fe] space, determined solely by the depletion rate of Eu from the ISM.}\label{F:RandomYields}
		\end{figure}
		This could be leveraged to provid a much greater constraint on the success of our models - the models of  Fig. \ref{F:Synthesis} undergo their collapsar cutoff during the thick/thin disc transition, and so we would expect to see the resultant radial patterns strongly rule out these model, even if properly calibrated. The restrictions imposed on our \modelname{} instantiations are far from the strictest we can generate, further emphasising that our conclusions are the upper bound of collapsar contributions. 
	
	\insertBigTable{}
	
	\subsection{Impact of the Simple Yield Approximation}\label{S:YieldApprox}

			In the absence of any observational constraints on the collapsar yield function $Y_\text{coll}$ we follow the literature (i.e. \siegel) by adopting the simplest possible form: that of \eref{E:Suppressor}. One could argue that this functional form is inappropriate, or otherwise corrupts any conclusions we might draw from our models. However, several factors justify that approach.
			
			The behaviour of Figs. \ref{F:MultiSACEM} and \ref{F:RamicesBacktrack} shows that even if the yield functions are not well-approximated by a simple function, the overall range in [Eu/Fe] is a necessity that cannot be mended by introducing an additional variation into $Y_\text{coll}$. Once collapsars stop forming, the [Eu/Fe] ratio drops at a fixed rate which depends only on the gas depletion timescale, and independent of any chosen form of the yield function. The only way to ensure that [Eu/Fe] $\approx 0$ at simulation end is by adding more Eu before collapsars die off. However, because collapsars are only active for a short time, no matter the functional form of the yield, increasing the Eu abundance pushes [Eu/Fe] above our `reasonable domain', and the model would be ruled a failure. We demonstrate this principle in Fig. \ref{F:RandomYields}, in which we replace our simple yield function with a positive random-walk function $y_{i+1} = |y_i + R|$ where $R$ is a random number in the range [-1,1], allowing for arbitrary functional forms of $y_\text{coll}(Z)$. We see that no matter the functional form, after the collapsar suppresion begins, all of the yields converge rapidly to a single path through [Fe/H]-[Eu/Fe] space. This path is determined solely by the required amount of Eu in the galaxy at the time that collapsars are suppressed, and hence is independent of the collapsar yield function: i.e., the models care how much Eu there is in the galaxy at $\tc$, not how it got there.
			
			The post-turnoff behaviour of these models would therefore be very well represented by a `simple'-type model which was tuned to produce the same Eu mass at collapsar turnoff as the varying model. For a given set of model parameters if a simple-type yield would cause the model to be ruled incompatible with the observed evidence at any time after collapsar suppression, then so would the corresponding model with a varying yield function. 
			
			By eliminating model classes based on their simple-yield approximations, we cannot be falsely eliminating any physically meaningful models, so we conclude that our results are robust against the impacts of the approximation of the simple form of $y_\text{coll}$.
	
	\subsection{Expanding the Search}
	
		In these initial forays we were altering a single parameter in a single `best fit' model, so one could argue that the search missed the right combination of parameters that allows for a high collapsar contribution despite a reasonable cut-off metallicity. 
		
		The challenge is that such a set of parameters must not simultaneously make, for example, the [Mg/Fe] histories unrealistic: we must ensure that any alterations produce a simultaneously realistic model in multiple chemical planes.

		\subsubsection{`Success` \& Viable Models}\label{S:Success}

		Following up on our summary in \S\ref{S:ObservationalSummary} we now define the criteria for a viable chemical evolution model. We emphasise that, in line with our efforts to falsify model classes rather than find a best fit model, the criteria developed here do not imply that a model is fully physical if it fulfils the criteria, but models that breach the criteria are clearly in contradiction to the empirical evidence.
		
		We define a galactic chemical history as `viable' if the resulting Tinsley curves for [Mg/Fe], [Eu/Fe] and [Eu/Mg] lie between the two black two curves in the corresponding panels of Fig.\ref{F:SAGAData}. Whilst a good model should also reproduce the observed distributions not just pass within its range, in keeping with our approach of negative inference, we instead choose to have room to be generous and yet still constrain the models. All three planes must simultaneously meet this criterion for the model to count as successful. 
		
		We make the intentional choice that the constraints only apply for [Fe/H] $>-1.5$. Though this weakens our constraints, it gives us numerous advantages: we limit ourselves to the strictly non-stochastic regime, avoid halo contamination and as per \ref{S:YieldApprox}, we avoid eliminating classes of models which have high variability at low metallicities.
		
		\newcommand\meanGrad{ \left\langle \frac{\d \text{[Eu/Fe]}}{\d t} \right\rangle }
		\newcommand\grad[1]{$\left|\meanGrad\right| \leq #1$ dex Gyr$^{-1}$}
		
		In addition, we study the effect of introducing a constraint on the time evolution of the models in concordance with the observational constraint in Fig.\ref{F:TemporalData}. Such Gradient-Constrained models are only considered successful if:
		\begin{equation}
		\left|\meanGrad\right| \leq 0.01 \text{ dex Gyr}^{-1}\label{E:GradientConstraint}
		\end{equation}
		where this is measured over the final 2 Gyrs of the simulation, and is in agreement with Fig. \ref{F:TemporalData}.
		
		Finally, a model galaxy must be capable of a sustained rate of star formation, even at late times. In GCE models it is common to constrain this through the star formation efficiency, represented by (up to a factor of order $\nu_\text{sfr}$) the quantity $M_\text{c}/ M_\text{*}$. For the Milky Way this value can be measured (i.e., \citealt{McKee2015}), indicating that we should constrain this value to  $\sim 0.1$. 
		
		We reemphasise that we have left the acceptance criteria intentionally generous: we want to find which models are excluded by failing to meet even these lax criteria.

	\subsection{Variable Selection for \modelname{}}\label{S:SACEMSearch}
		
		We perform a Monte-Carlo exploration of the high-dimension parameter space by randomly and independently selecting the model parameters (except $\chi$, the collapsar contribution ratio, and $\tc$, the collapsar cutoff time) from between bounds determined by the imposed constraints (see Table \ref{T:Parameters}). For each random set of parameters, we then perform a sweep of $\chi$-$\tc$ space, such that each instantiation is evaluated uniformly across this space. Note that the enforcement of the temporal gradient constraint (Eq. \ref{E:GradientConstraint}) plays a special role, as we perform all simulation runs both with and without this constraint. 
		
		We call the different constraint sets in Table \ref{T:Parameters} the \texttt{Unconstrained} set (denoted \texttt{U}), \texttt{Weak} (\texttt{W}) and \texttt{Viable} (\texttt{V}). As indicated by the name, in Set \texttt{U}, parameters are allowed to take almost any value, with essentially no constraint or regard for their physical reality. Naturally, the `unconstrained' set is not truly unconstrained. Instead we mean that the boundaries imposed reflect our desire to search  `interesting' regions of parameter space within finite computation time, and do not meaningfully eliminate any kinds of physically interesting galaxies we might care to consider. We include this set to act as a null hypothesis - that without physical constraints, models can be made to fit to the data - and hence that any region of parameter space that is eliminated is due to the imposition of physical constraints. 
		
		The \texttt{Weak} set of parameters (\texttt{W}) adds an amount of physical intuition - the order of magnitude of the galactic mass is determined, cooling timescales are of an order close to what we might expect, and so on. These constraints encapsulate the behaviour of most galaxies, but do not uniquely identify the Milky Way.  In contrast, the \texttt{Viable} set imposes the minimum requirements to match the properties of the Milky Way. 
		
		A fourth variant, the \texttt{Mixed} model (\texttt{M}), was also studied, which uses the \texttt{Viable} constraints for all parameters except those relating to the SFR, which are treated as \texttt{Unconstrained}. The mixed set forms the basis of the discussion in \S\ref{S:Properties}.

\section{Results}\label{S:Results}

	\begin{figure*}
	    \centering
	    \fittedImageLarge{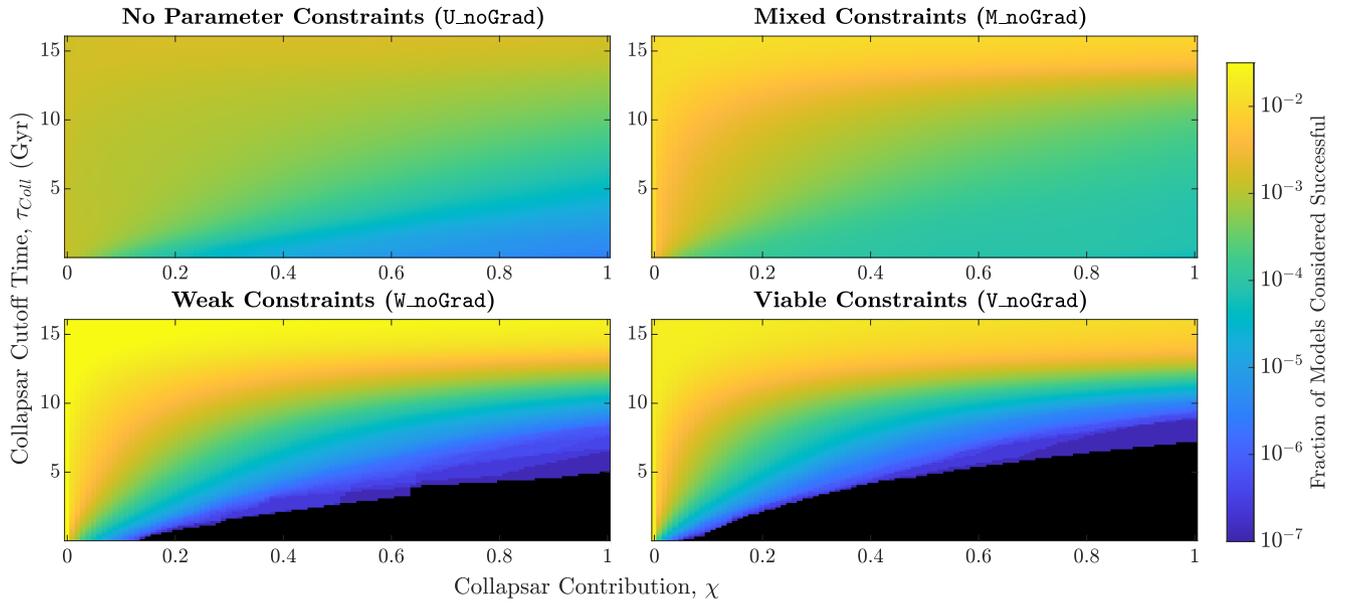}
	    \caption{The results of the \modelname{} gridsearch for models where no gradient check was performed, and where the fraction of europium, $\chi$, was measured from the present-day cold gas reservoir. Black regions are those in which no successful models (as determined by the success conditions of \S\ref{S:Success}) could be found. Note that the $\tc$ axis does not extend exactly to $\tc =0$ (the region where collapsars are never active), as $\chi \neq 0$ is nonsensical if collapsars are inactive for all history. Therefore $\tc \geq 0.16$Gyr.}
	    \label{F:Cold0}
	\end{figure*}
	
	\subsection{\modelname{} Search}\label{S:SearchResults}

		For each of the constraint sets of \S\ref{S:SACEMSearch} we randomly generated $10^7$ parameter sets, which were then each evaluated across a 101$\times$101 grid in $\chi-\tc$ space, for a total of $\sim 10^{11}$ models.

		Due to the high density of generated models, we may therefore be confident that the regions in which no models could be found (the `exclusion zone') are genuine forbidden regions of parameter space. This inference is strengthened if the exclusion zone are contiguous: whilst statistical fluctuations may alter the position of the boundary, the contiguous nature implies that regions a significant distance from this boundary are robustly excluded.

		\subsubsection{Without Temporal Gradient Constraints}
		
			We first consider those models which did not use the condition of \eref{E:GradientConstraint} to evaluate the success of a model. The density of successful models is shown in Fig. \ref{F:Cold0}, with regions coloured black denoting the points where no successful model could be found. 
			
			As expected, for the \texttt{Unconstrained (U)} and \texttt{Mixed (M)} constraints, we are able to find allowed models for all values of the collapsar contribution $\chi$ and the cutoff time $\tc$. However, the density variation for models with large-$\chi$ but low $\tc$ (i.e. lower right of each panel) shows that such models are disfavoured even with these lax constraints:  in both the \texttt{Unconstrained} and \texttt{Mixed} investigations, NSM-dominated models (left of each panel) were favoured by a factor $\sim 500$ over the corresponding high-$\chi$-low-$\tc$ models.

			With the introduction of the only \texttt{Weak} constraints, we see the formation of an exclusion zone in the high-$\chi$, low-$\tc$ region: the \texttt{Weak} constraints fail to find any models which are collapsar dominated ($\Omega \sim 1$) for $\tc < 6$Gyr, and limits the Milky Way to $\Omega > 0.65$ for $\tc < 3$Gyr. The exclusion zone is not qualitatively changed by the introduction of the \texttt{Viable} constraints, but the size increases such that collapsar dominated models are prohibited for $\tc < 7$ Gyr, $\Omega > 0.5$ is prohibited for $\tc < 4$Gyr and $\Omega > 0.2$ is prohibted for $\tc < 2$Gyr.

			This immediately brings us into direct tension with the theoretical models of collapsars. As per \S\ref{S:Cutoff}, no theoretical models yet allow for events above $Z = 0.3 Z_\odot$, but following the discussion in \S\ref{S:MetalDecouple} and the work of i.e. \cite{Casagrande2011}, we know the Milky Way was rapidly enriched early in its lifetime. In other words, $\tau \sim 2$Gyr is the approximate time coordinate that we should associate with the `maximum theoretical collapsar time'. 
			
			These initial results therefore strongly indicate that, under \texttt{Weak} and \texttt{Viable} constraints, the r-process enrichment of the Milky Way cannot be dominated by collapsars events constrained to occur before $Z = 0.3 Z_\odot$. 
			
			However, in \S\ref{S:MetalDecouple}, we cautioned against inferences which rely on coupling $\tc$ and $\Zc$ too tightly, as there is a small amount of flexibility in the relationship due to our decoupling of the background metallicity evolution from the properties of galaxies. The \texttt{Viable} exclusion zone in Fig. \ref{F:Cold0} has a boundary at $\sim 4$Gyr, so between the Monte Carlo search and the decoupling of $\Zc$ and $\tc$, it is plausible that our results might still allow a collapsar dominated galaxy which we failed to detect: more work is yet needed.

			\begin{figure*}
			    \centering
			    \fittedImageLarge{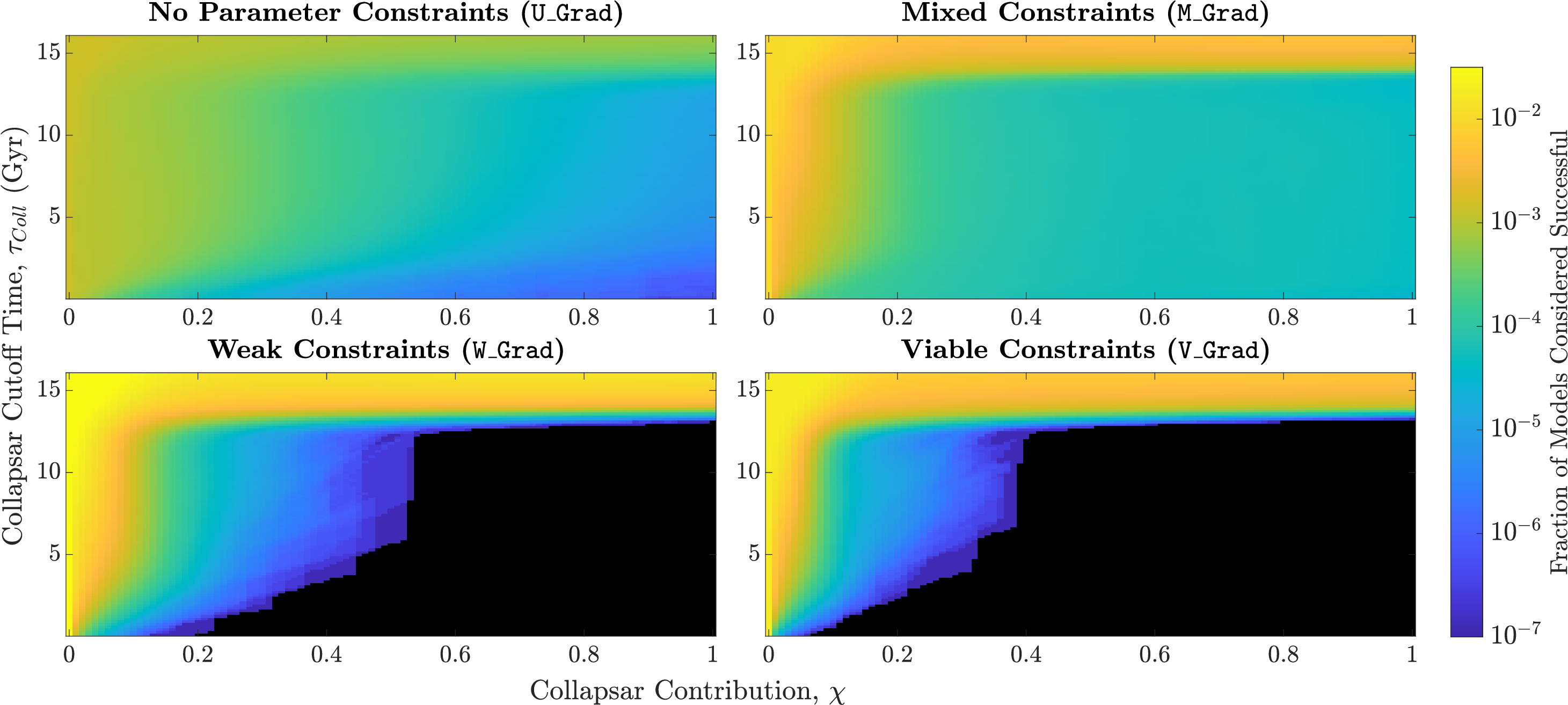}
			    \caption{As with Fig. \ref{F:Cold0}, with the inclusion of the constraint that only models with \grad{0.01} can be considered successful. }
			    \label{F:MaximumConstraint}
			\end{figure*}
		
		\subsubsection{Temporally Constrained Models}
		
			In Fig. \ref{F:MaximumConstraint} we examine the same set of constraints as Fig. \ref{F:Cold0}, but with the additional constraint that successful models must obey \grad{0.01} across the final 2Gyr of evolution. We note from Fig. \ref{F:TemporalData} that this is still a generous constraint. 
			
			We see that, once more, models which use \texttt{Unconstrained} and \texttt{Mixed} constraints excluded no regions of parameter space, though as before they favoured low-$\chi$ or high-$\tc$ models by more than a factor of $10^2$ and $10^3$ respectively. 
			
			However, for \texttt{Weak} and \texttt{Viable} models, we see that large regions of space have been declared non-viable. Comparing models \texttt{W\_noGrad} and \texttt{W\_Grad} (the lower left panels of Figs. \ref{F:Cold0} and \ref{F:MaximumConstraint} respectively) we see that the addition of the temporal gradient constraint has expanded the exclusion zone. The region $\chi > 0.6$ (previously excluded for $\tc < $6Gyr) is now excluded for all times $\tc < 12$Gyr. This makes the exclusion zone in this region extremely resistant to both statistical errors and errors arising from the imposed metallicity function, as such errors would need to cause errors on the order of 5Gyr, which we do not consider a reasonable expectation.  
			
			Similarly, with the addition of the temporal constraints, all successful \texttt{Viable} models were limited to $\chi < 0.4$ for $\tc < 12$Gyr and $\chi < 0.3$ for $\tc < 4$Gyr. This eliminates all models for which collapsars are either the majority or dominant source of r-process nucleosynthesis, as there is substantial theoretical and observational evidence that collapsars cannot have been occuring more recently than 2Gyr ago. 
			
			Again, we emphasise that the vertical expansion of the exclusion zone observed in models \texttt{W\_Grad} and \texttt{V\_Grad} makes these conclusions extremely robust to statistical and $Z_c-\tc$ induced errors. We are therefore condfident that these regions are truly excluded portions of parameter space.

	\subsection{\codename{} Search}
	
		We performed a similar search on the \codename{} simulation, though due to computational limitations, the number of models is much smaller. In addition, \codename{} is more physically motivated than the semi-empirical approach of \modelname{}, such that there are fewer parameters which we may alter without undermining some other observable property of our galaxy. In particular, the SFR (which is coupled to gas infalls and radial gas motion) is chosen to replicate features of the solar neighbourhood (see \sbRef). The remaining free parameters which are not fixed to replicate observable properties of the galaxy are therefore:
		
		\begin{itemize}
			\item The hot-gas injection fraction $f_\text{CCSN}$ and $f_\text{NSM}$
			\item The delay time of SNIa processes, and the fraction of white dwarf remnants which can undergo SNIa events
			\item The fraction of synthesised material lost to the IGM
			\item The effective r-process yields $\bar{Y}_\text{Eu}^\text{s-process}$, $\bar{Y}_\text{Eu}^\text{NSM}$ and $\bar{Y}_\text{Eu}^\text{Coll}$
			\item The value and width of the metallicity cutoff.  
		\end{itemize}
		
		Because the \codename{} model is inherently much more tightly constrained, we widened the success condition to prevent overfitting of the model. The only success condition applied to the \codename{} models is that, for the solar radius, the final [Eu/Fe] value must lie within the `viable domain' of Fig. \ref{F:SAGAData} (the final [Fe/H] value already being calibrated to the solar neighbourhood).
		
		Note that, unlike \modelname{}, this simulation does use metallicity for the collapsar cutoff, rather than a metallicity-inferred time. We also note that due to the properties of the \codename{} simulation, we cannot target a specific final collapsar-contribution fraction $\chi$; we have to generate models with a given set of europium yields and determine $\chi$ at simulation end. Hence, unlike the models of the previous section, the models are not produced on a uniform grid of $\chi$, and due to the computational constraints it proved somewhat difficult to even generate a meaningful number of models with $\chi \sim 1$ at low $\Zc$. In particular, we were not able to generate a single model for which $\Zc = 0.1Z_\odot$ but $\chi > 0.85$. However, given the contiguous (and large) nature of the final exclusion zone, we do not expect this to impact our final conclusion.
		
\begin{figure}
		    \centering
		    \fittedImage{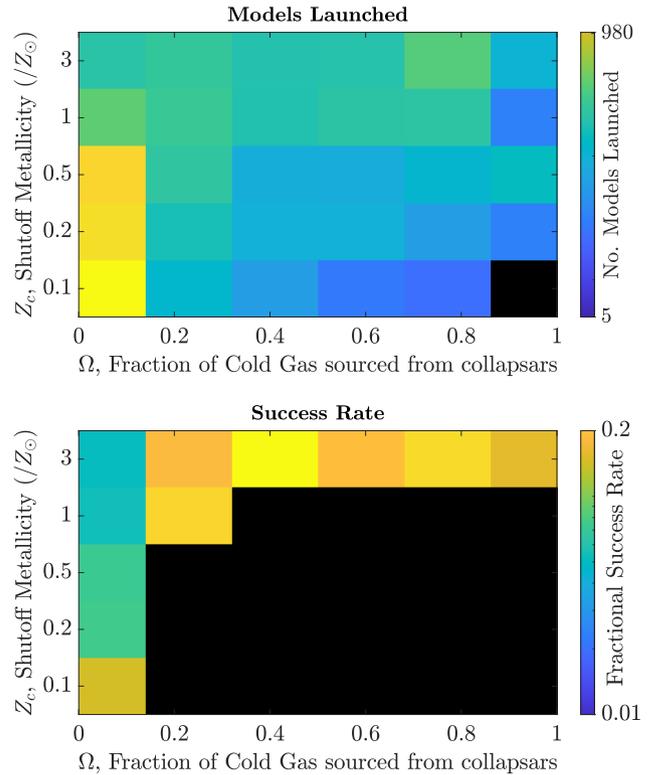}
		    \caption{The equivalent success plot to those shown in figures \ref{F:Cold0}-\ref{F:MaximumConstraint}, generated instead from the \codename{} simulation. The top panel shows the density of all models which were launched (\codename{} measures $\chi$ at simulation end). The lower panel shows the ratio of successful models to those launched.}
		    \label{F:RamicesGrid}
		\end{figure}

		We ran 3634 iterations of the \codename{} simulation, with the subset of variables drawn from the \texttt{Weak} sample shown in Table \ref{T:Parameters}. The results of this search are plotted in Fig. \ref{F:RamicesGrid}. Despite the fact that the random parameters were generated with the \texttt{Weak} model, with no explicitly included gradient consideration and with an even more generous acceptance criteria, \codename{} produced a grid with an even larger exclusion zone than the temporally-contrained \texttt{Viable} model of \modelname{}.
		
		It seems that the initial results we inferred from Fig. \ref{F:RamicesBacktrack} -- that a multi-zone model with more physically motivated parameters suffers a significantly larger drop in [Eu/Fe] than the corresponding \modelname{} model -- continues to hold, thereby indicating that the \modelname{} results should be interpreted as an upper bound on the maximum collapsar contribution.

\section{Properties of Successful Models} \label{S:Properties}

	Figs. \ref{F:Cold0} and \ref{F:MaximumConstraint} demonstrate that the metallicity dependence of collapsar events necessarily limits them to sub-dominant contributors to the galactic r-process budget, yet it is more instructive to exmaine in more detail those models deemed ``successful''.
	
	\modelname{} is unique in its low computational footprint, so we can for the first time systematically scan the full parameter space. This allowed us to entertain non-physical sets of parameters in our search for the minimum required set of constraints: the models generated from the \texttt{Unconstrained} and \texttt{Mixed} sets, in particular. 
	
	This was, primarily, an aesthetic choice: we wished to impose as few additional constraints onto our models as possible. However, by examining the behaviour of the global parameters from the less-constrained models within the  $\Omega-\tc$ plane, we may infer how the successful models fulfill the imposed requirements. This provides three benefits: i) comparing how past studies have been able to match chemical observations, and if this lies in tension with other observables ii) examinations into the limitation to these approaches, i.e. how far we may `tweak' the parameters of our model whilst remaining physically viable and iii) examining if there are other ways for a model to fit the data beyond the canonical understanding. 
	
	\subsection{Star Formation}
	
		\begin{figure}
		    \centering
		    \fittedImage{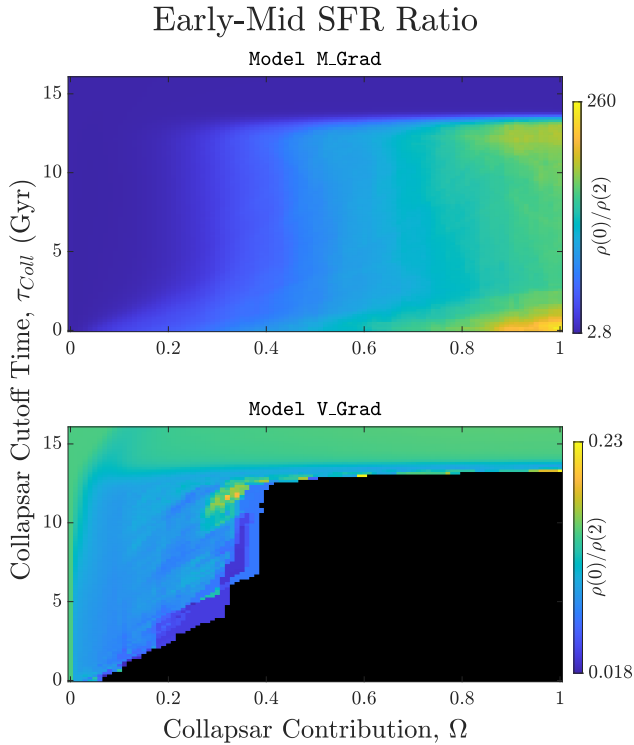}
		    \caption{The mean ratio $\rsfr(0)/\rsfr(2$Gyr$)$, a measure of the `peakiness' of the early time Star Formation Rate as measured across the \texttt{M\_Grad} and \texttt{V\_Grad} models.}
		    \label{F:SFR}
		\end{figure}
		
		\begin{figure}
		    \centering
		    \fittedImage{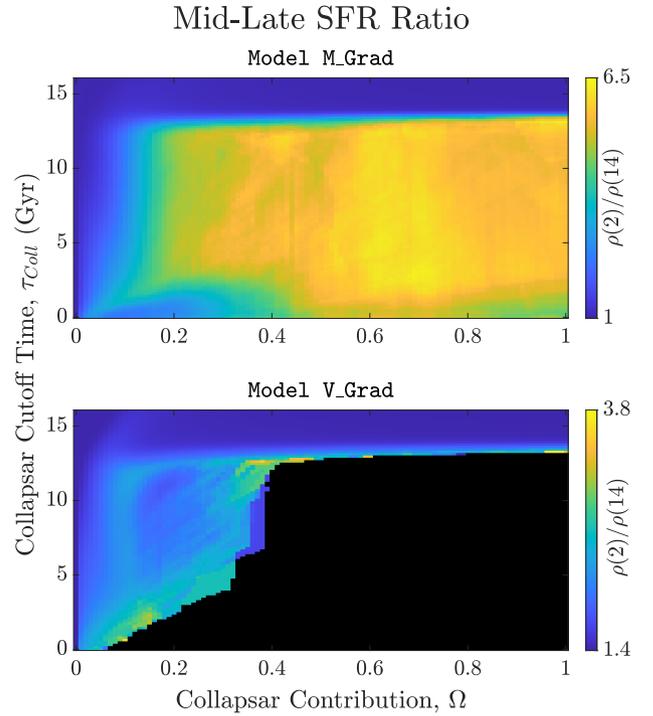}
		    \caption{As with Fig.\ref{F:SFR} but with $\rsfr(2)/\rsfr(14)$: a measure of how quickly the star formation in the galaxy dies down.}
		    \label{F:SFR_End}
		\end{figure}
		
		The most straightforward way for a model to produce a required chemical history within the specified bounds is with an extremely tightly controlled star-formation rate. Within \modelname{}, if a model has a highly-peaked early-time SFR, then the large collapsar population could generate a vast amount of Eu over a short period of time. In such models, a true GCE model would also accumulate an equally vast number of metals during this period due to the high SFR, such that collapsars should shut off. However, because the imposed function $\Z(t)$ is insensitive to the particular parameters of our model, the `true metallicity' of the model diverges significantly from the nominally assumed model, and so collapsars will continue contributing long past the time they should have died out. 
		
		This therefore enables models with even tiny values of $\tc$ to be considered viable, where a more physically coupled SFR and metallicity evolution would discard these models. We suspect that the vast majority of $\tc-\chi$ space accepted under the \texttt{Mixed} regime, but rejected under \texttt{Weak} and \texttt{Viable} models are achieving their success through this method. 
		
		Figs. \ref{F:SFR} and \ref{F:SFR_End} justify this claim, they respectively depict the ratio of the initial star formation rate and after 2Gyr: $\rsfr(0)/\rsfr(2\text{Gyr})$ and the corresponding ratio between 2Gyr and the end of the simulation, $\rsfr(2\text{Gyr})/\rsfr(14\text{Gyr})$, for two classes of models. The properties of the models in the high-$\chi$-low-$\tc$ region of \texttt{M\_Grad} are striking: the search preferentially found models where the initial SFR was more than 2 orders of magnitude greater at simulation start than it was just a few gigayears later, and where the mid-time SFR was on average a factor of 6 greater than the SFR at simulation end, such that there is a total of more than 3 orders of magnitude between the $t = 0$ SFR and the final time SFR.  These ratios are orders of magnitude outside the constraints measured for the star formation history of the MW, which typically show a decline in SFR by less or equal to an order of magnitude \citep{Mor2019,Schonrich20091, Aumer2009}. 
		
		By comparison, the \texttt{V\_Grad} model shows no particular bias in either of Figs. \ref{F:SFR} or \ref{F:SFR_End}, indicating that large-scale constraints such as the total mass of the Milky Way are reasonable methods to eliminate these models.

	\subsection{Hot Gas} 
		
		The inclusion of a thermal gas phase is a relatively rare feature of GCE models, though recent work has highlighted its importance \citep{Khoperskov2021, Schonrich2019}. Within \modelname{} and \codename{} we implement the thermal phase in a relatively crude fashion: a distinct hot and cold phase, with the hot phase being fed with a fraction $f$ of the enriched gas, and then `cooling' into the cold phase with a decay frequency $\lambda$ such that $\dot{M}_\text{cool} = \lambda M_\text{hot}$.
		
		\begin{figure}
		    \centering
		    \fittedImage{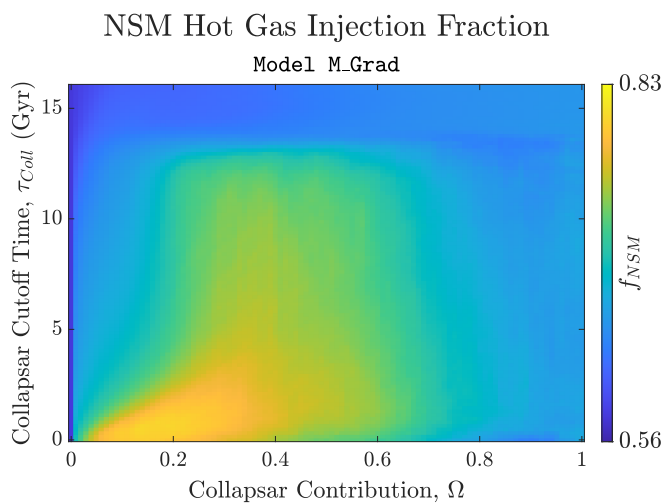}
		    \caption{The mean value of $f_\text{NSM}$, the fraction of NSM ejecta placed into the hot gas reservoir for successful \texttt{M\_Grad} models.}
		    \label{F:NSMInjection}
		\end{figure}
		
		Fig. \ref{F:NSMInjection} shows the variation of the parameter $f_\text{NSM}$ in the successful models, which displays an interesting pattern, which we argue both vindicates the inclusion of the hot phase, and is central to our heuristic understanding of the interplay between NSM events and collapsars. 
		
		At very low $\Omega$ ($\ll 0.1$), this value is favoured to be low, around 0.6 or below (for comparison, the corresponding CCSN, Collapsar and SNIa values exceed 0.9 almost everywhere) - this was argued for in \cite{Schonrich2019} as a way for the thermal properties to permit NSM to be viable contributors at early times, due to the correspondingly higher immediate contribution to the star forming reservoir. As collapsars become more prominent at early times (higher $\Omega$ and lower $\tc$), it is no longer necessary to invoke the arguments of SW19 to explain the early time [Eu/Fe] values, and hence the need for low values of $f_\text{NSM}$ to meet the chemical criteria at [Fe/H] = -1.5 vanishes, leading to the observed increase in $f_\text{NSM}$. At even higher values of $\Omega$, the value of $f_\text{NSM}$ decreases again - notably to a value of 0.65, which is the exact midpoint of the permitted range: hence we are seeing ambivalence to the value of $f_\text{NSM}$ when determining if a model is successful. However, for $\Omega < 0.7$ there are clear signs that the thermal properties of NSM are very important for determining the success or failure of a model.

\begin{figure}
		    \centering
		    \fittedImage{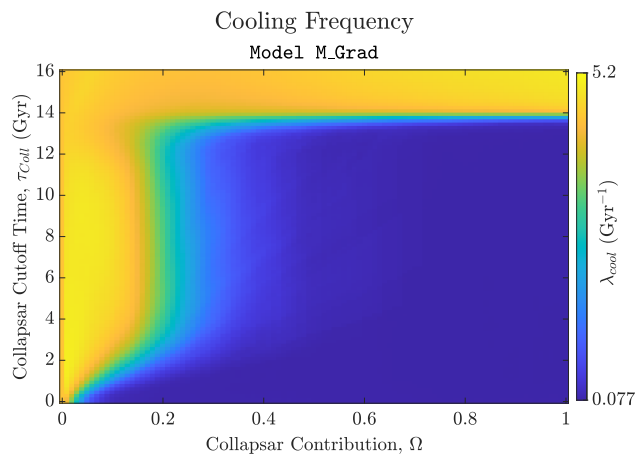}
		    \caption{The mean ratio $\lambda_\text{cool}$, the hot gas cooling rate, for sucessful models constrained with \texttt{M\_Grad}. Note that the midpoint of the permitted range for $\lambda_\text{cool}$ is 5 - models close to this value are likely to be insensitive to the value of $\lambda$. }
		    \label{F:Cooling}
		\end{figure}

		Given that the thermal properties have been shown to be important, we now consider the case where the cooling rate is small. If the hot-gas injection fractions are non trivial, then a large portion of the enriched gas can be secreted away in the hot gas phase, preventing over-enrichment at early times, and providing a `source' of enriched gas into the star-forming phase long after the gas itself was actually produced. Whilst this is desirable to an extent - it is reasonable to expect exactly this to happen - Fig. \ref{F:Cooling} shows that the successful models with a large collapsar contribution were almost overwhelmingly those which abused this property to the extreme. 
		
		In Fig. \ref{F:Cooling} we see that for $\chi > 0.3$ and $\tc < 13$Gyr, the mean value of $\lambda$ is $\approx 0.077$ Gyr$^{-1}$, an order of magnitude below the commonly used value of $\sim 1$ Gyr$^{-1}$. These models cool almost no gas into the cold gas phase, allowing the hot gas phase to become hyper-enriched relative to the cold gas phase. This shows that these successful models favoured extreme thermal fractionation as an explanation for the chemical history of the galaxy - whilst we do not doubt that a hot gas population is important for understanding GCE, the cooling rates indicated here strongly suggest that these models are highly unphysical.

	\subsection{Lockup Modification}
	
		\begin{figure}
		    \centering
		    \fittedImage{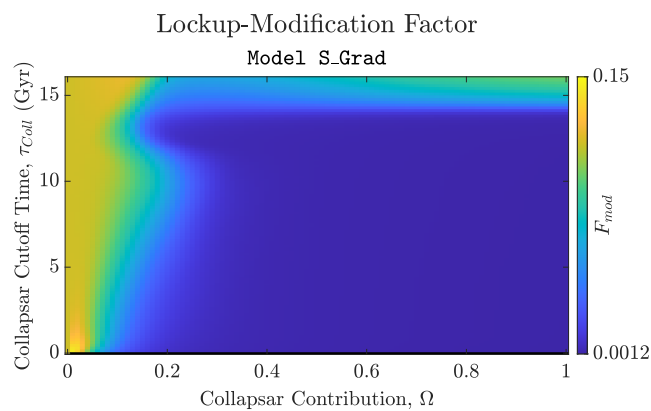}
		    \caption{The mean value of $F_\text{mod}$, the lockup-modification factor, for successful models at each point in $\tc-\chi$ space for the prototype simulations in which $F_\text{mod}$ is bounded by $0 < F_\text{mod} \leq 1$. We note that the black strip at $\tc = 0$ is the nonsensical combination $\tc = 0, \chi > 0$. As per Fig. \ref{F:Cold0} this was omitted it all other plots.}
		    \label{F:Lockup}
		\end{figure}
		
		The lockup modification factor, $F_\text{mod}$, encodes the rate at which synthesised material is removed from the cold gas reservoir by star formation, such that the lockup rate is $\propto F_\text{mod}\rho_\text{sfr}$ rather than the zeroth-order approximation $\propto \rho_\text{sfr}$. $F_\text{mod}$ differs from 1 since, though normal CCSN events may not synthesise new Eu material, they can recycle pollutant metals back to the ISM, thereby reducing the effective lockup rate of synthesised material. 
		
		We must be aware, however, of the extreme case $F_\text{mod} \approx 0$ in which no synthesised material can be locked up. Star formation would continue to deplete the unenriched gas mass, such that the stars are preferentially forming from primordial material: the cold gas phase becomes chemically fractionated, and becomes ever more enriched without any additional synthesis events. 
		
		Collapsar models with $\tc < 4$Gyr could exploit this property to `hide' a reservoir of Eu which persists until late times, when by all physical reasoning, it should have been depleted by the lockup of the continual star formation. This was made evident in a prototype set of simulations (termed the \texttt{Old-Style} simulations), in which the lower bound on $F_\text{mod}$ was set to be $0$ for all simulations. Fig. \ref{F:Lockup} shows the behabiour of $F_\text{mod}$ of successful models in the prototype simulation. We note that the prototype simulations differ from the final ones in several ways that make a direct comparison difficult, but we can see that the lower right section of the Fig. \ref{F:Lockup} is overwhelmingly dominated by models for which $F_\text{mod} \sim 0.001$, the case where there is no chemical lockup. A similar pattern was observed for models \texttt{U, W} and \texttt{V}: an overwhelming bias towards extremely small values of $F_\text{mod}$ in the region of high-$\chi$ and low-$\tc$. 
		
		Such models are evidently unphysical, and so in the final suite of simulations, we bounded $F_\text{mod}$ from below by 0.3. This value was motivated by comparison with a reasonable IMF, noting that $\int_0^{1} M \zeta(M)\d M > 0.35$ for all commonly used IMFs \citep{Salpeter1955,Chabrier2003}. Since stars with $M = M_\odot$ have lifetimes $\sim$10Gyrs, at least this fraction of the gas must be locked-up for long timescales.

		When bounded in this fashion, the bias in the values of $F_\text{mod}$ vanished almost entirely, and so we surmise that we have closed off this unphysical route to achieving `success'. We note that some values near the boundary of the exclusion zone of both \texttt{W\_Grad} and \texttt{V\_Grad} did show a small bias towards smaller values of $F_\text{mod}$, indicating that if we were to impose stricter and more physical constraints on the value of this parameter, these models would similarly be eliminated. However, in the spirit of our attempt to find the minimum possible set of constraints, we leave these potentially problematic models unchallenged, as the size of the exclusion zone is already sufficient to draw our conclusions.   	
	
	\subsection{Improperly Coupled Models}
	
		In the above discussion, we saw a number of traps which models can fall into: failing to properly lock up their materials, poor treatment of hot gas phases and associated cooling rates and torturing the SFR until it allows you to replicate your desired features. In utilising these `traps', the models seemingly satisfy all chemical constraints which were placed upon them. However, on closer inspection these models were only able to reproduce the chemical properties due to an unphysical assumption elsewhere in the model.  
		
		The general theme of these assumptions was that they allowed the chemical reservoirs to become separated or fractionated in some way, such that the evolution of the SFR, cooling, chemical enrichment and subsequent lockup of the reservoirs was not functioning properly. 
		
		We note, for example, that \siegel{} use an SFR which is decoupled from the present gaseous or stellar mass in their model, which we suggest falls into a similar camp of improperly coupled models, and explains why their findings are in such strong tension to our own, despite seeming to meet all of the observable chemical criteria. 
		
		We also suggest that, due to the expansive sampling of parameter space, we would have been able to notice if some unusual combination of physical properties were able to replicate both the chemical data, and not fall into one of the four `traps' outlined above - in fact, we observed no such signal. Given that this is the case, it must be true that any successful GCE model which replicates a collapsar-dominated galaxy must strongly deviate from the physics encapsualted in the core equations of Appendix \ref{A:Model}.

\section{Conclusions} \label{S:Conclusions}

	In this paper, we have performed the first comprehensive chemical evolution study which examines the multi-dimensional parameter-space associated with the origin and evolution of galactic r-process material. In this extensive analysis, we could find no viable model with collapsars as the dominant source for today's r-process element budget.

	 In this work, we have introduced our newly developed Simple Analytical Chemical Evolution Model (\modelname{}). \modelname{} is an analytical framework, which incorporates the relevant physics (star formation histories, inflow, outflow of gas, yields to both a hot and cold gas phase, cooling of material from the hot phase, star-forming ISM, and different temporal and thermal properties for different sources of yields), but at the same time has run-times of fractions of a second, i.e. orders of magnitude faster than existing chemical evolution codes. Although \modelname{} relies heavily on some simplifying approximations (namely single-zone space with instantaneous mixing, empirical fixing of yields, and does not consider the lifetime of stellar populations), we found good agreement with the full chemodynamical models from \cite{Schonrich2009}, which does not make such approximations. 
	 
	 Where there are divergences between the models, we found that \modelname{} was more generous to collapsar-dominated models than the \sbRef work, though we note that both \modelname{} and \codename{} do not directly consider tertiary sources of r-process material (such as the i-process or magneto-rotational supernovae), the impact of dwarf galaxy accretion, and assume a simple form of the collapsar yields. Although these omissions might limit the generality of our results, we have justified their long-term impacts on the abundance patterns as negligible, or already encapsulated in part by features of our models. Hence, our results are robust against these approximations.

	A central problem holding chemical evolution studies back has been a reliance on costly models in a high-dimensional parameter space, which has forced prior studies to operate with exploratory modelling of a small number of models. \modelname{}'s performance allowed us to run $>10^{11}$ models, mapping out the full parameter space of r-process chemical evolution with both collapsars and neutron star mergers, and allowing us to pursue an entirely different strategy: instead of trying to find models that match some observational constraints, we drew up a full set of ``minimal consensus" observational constraints which models must replicate, and look for those models which fail to reach even these lax conditions: rough `bounding boxes' that our chemical tracks in [X/Fe]-[Fe/H] must pass through (Fig. \ref{F:SAGAData}), chemical equilibrium imposed across the final 2Gyr of evolution, and a sustained rate of star formation. Our search through parameter space was bounded by imposed conditions on the allowed parameter values -- the one which best represented the Milky Way (whilst still allowing for large degrees of variation) we termed the \texttt{Viable} set of constraints. We also explored the parameter space with unphysically lax constraints on parameter values (the \texttt{Unconstrained} and \texttt{Mixed} sets) - comparing these with the \texttt{Viable} constraint set reveals and analysis how classes of models with dominant r-process contributions discussed in the literature appeared to satisfy observational constraints. However, with the \texttt{Viable} constraints, we have found that:

	\begin{enumerate}
		\item no \modelname{} model could be found where collapsars contribute more than 30 per cent of the modern r-process budget, as long as collapsars were suppressed as in \cite{MacFadyen1999}. Neutron Star Mergers were always required to be dominant (Fig. \ref{F:MaximumConstraint}).

		\item A significant collapsar contribution at early times was not eliminated: Many models in which NSM are responsible for > 99 per cent the Eu abundance at late times had > 50 per cent collapsar contributions at $t < 1$Gyr (Fig. \ref{F:Synthesis}).

		\item The \codename{} code shows that the remaining parameter space allowed by \modelname{} still contains models that are in stark contradiction with the data. In particular the metallicity dependent cut-off can introduce a radial [Eu/Fe] increase in the galactic disc (Fig. \ref{F:RadialPlot}), which starkly contrasts with observations: our limit of <30 per cent is likely still too high, and can be refined further.
	\end{enumerate}

	We deliberately chose constraints which were overly generous, and our results should be seen as the maximum possible contribution of collapsars to the modern r-process budget. We leave further discussion regarding how far the constraints can be pushed and how far the allowed parameter space can be further shrunk to future studies, preferring to keep our argument simple: we have shown that even a minimal set of constraints permits no models with collapsars as dominant source of r-process elements, and thus leaving by exclusion \citep{Sneden2008} only neutron star mergers as a dominant source.

	\section{Acknowledgements}
	
	We thank the Oxford galactic dynamics group for their helpful discussions, in particular J Binney for his extensive and rigorous comments. JF acknowledges the support from the and the Royal Society grant RGF\textbackslash EA\textbackslash 180290. RS is supported by a Royal Society University Research Fellowship. This work was performed using the Cambridge Service for Data Driven Discovery (CSD3), part of which is operated by the University of Cambridge Research Computing on behalf of the STFC DiRAC HPC Facility (www.dirac.ac.uk). The DiRAC component of CSD3 was funded by BEIS capital funding via STFC capital grants ST/P002307/1 and ST/R002452/1 and STFC operations grant ST/R00689X/1. DiRAC is part of the National e-Infrastructure
	
	\section{Data Availability}
	
	We anticipate making the \modelname{} code publicly available in a future publication. Until that time, the model and the corresponding data will be shared on request to the corresponding author. 
	
\bibliographystyle{mnras}
\bibliography{CollapsarCitations}

\appendix
\section{Europium as a tracer}\label{A:Tracer}
	
	It is convenient for us to use a proxy for the total r-process enrichment: Europium (Eu). Europium is chosen over similar r-process elements such as Gadolinium (Gd) and Dysprosium (Dy), since Europium is one of the purest r-process elements, being $\sim$98 per cent r-process in origin (\sneden{}). Europium has strong spectral lines in the optical spectrum, with well known oscillator strengths \citep{Biemont1982} and a large amount of associated data. 
		
	The usage of a direct proxy might be called into question due to hints of differing trends between Eu, Gd and Dy \citep{Guiglion2018}, and the breakdown of the `common fingerprint' for lighter r-process elements (see \sneden{}). Whilst the effects of the assumptions of Local Thermodynamic Equilibrium have been studied in Europium (i.e. \citealt{Zhao2016}), such corrections have not been calculated for Gd and Dy. It is therefore plausible that the reported differences in the trends between Eu, Dy and Gd vanish upon full consideration of NLTE and 3D atmospheric modelling. 
	
	On balance, Eu serves as a convenient proxy for r-process elements. 
	
\section{Understanding Fig. 1} \label{A:DataExplain}
	
	We present a brief description of how the standard models of GCE predict and explain the behaviour exhibited in Fig. \ref{F:SAGAData}, in particular the flat line ('plateau') at low metallicities, and the downturn ('knee') seen for [X/Fe] at [Fe/H] $\sim -1$. In the [Mg/Fe]-[Fe/H] plane, the plateau arises from pre-SNIa, CCSN-dominated enrichment: at early times the [Mg/Fe] ratio is dominated by the yield ratio from massive stars, i.e.:
	
	\begin{equation}
		\left[ \frac{\text{Mg}}{\text{Fe}} \right] \approx \frac{ \int_0^t \rsfr(t) \d t \int_{M_\text{min}(t)}^\infty \zeta(M) Y_\text{Mg}(M,Z(t)) \d M}{\int_0^t \rsfr(t) \d t \int_{M_\text{min}(t)}^\infty \zeta(M) Y_\text{Fe}(M,Z(t)) \d M} \label{E:InitialRatio_time}
	\end{equation}	
	
	Here  $\zeta(M)$ is the Initial Mass Function (IMF) and $Y_\text{X}(M,Z)$ is the net amount of element $X$ synthesised by a star with progenitor mass $M$ and metallicity $Z$. The integrals are bounded from below by $M_\text{min}(t)$, the minimum mass star which has reached the end of its lifetime at time $t$. Under the approximation that the variation in $M_\text{min}(t)$ and $Z(t)$ does not alter the value of the integral over short timescales, the time dependence drops out, leaving a constant abundance ratio at early times:
	
	 \begin{equation}
		\left[ \frac{\text{Mg}}{\text{Fe}} \right] \approx \frac{  \int_{M_\text{min}}^\infty \zeta(M) Y_\text{Mg}(M,0) \d M}{\int_{M_\text{min}}^\infty \zeta(M) Y_\text{Fe}(M,0) \d M} \label{E:InitialRatio}
	\end{equation}	
	
	For a similar discussion, see \cite{Weinberg2018}. As more stellar evolution is allowed to occur, SNIa events can kick in (initially being prohibited by longer progenitor lifetimes and subsequent inspiral or accretion phases). SNIa heavily favour the synthesis of iron-peak elements over $\alpha$ elements such as magnesium \citep{Iwamoto1999}, so we will see a decrease in all [X/Fe] planes where X does not have significant SNIa production - this is the `knee' seen at [Fe/H] $\sim$ -1. This simple outline (neglecting confounding factors such as the thermal phases of the ISM or metal loss from the galaxy) covers the main patterns seen in the chemical evolution of [Mg/Fe]. 
	
	Na\"ively, it is surprising that [Eu/Fe] in Fig. \ref{F:SAGAData} behaves similarly to the canonical picture of [Mg/Fe], as we do not expect CCSN to contribute significantly to Europium synthesis (indeed, this is why europium was chosen as a tracer). This has often been used as an argument in favour of Collapsars being the dominant r-process source.

\section{Analytical Model}\label{A:Model}
	
	\subsection{Star Formation}
		
		Our analytical model, \modelname{}, derives its time-dependent star formation rate from a physically motivated model of the galaxy and the accretion and heating of three gas reservoirs ($M_c$, the cold gas, $M_h$, the hot gas, and $M_*$ the mass locked up in stars). We use a Kennicut-Schmidt style model for star formation: stars form at a rate given by $\dot{M}_{*, born} = \nu_\text{SFR} M_c$, and we use a simple `exponential death' model for stars returning their material back to the ISM, such that $\dot{M}_{*, die} = \mu M_*$ (note that this is solely for the purposes of the SFR and does not alter the chemical evolution. A more complex prescription could be used, but the final result would be equally replicated by altering $\nu_\text{sfr}$ or $\mu$, and therefore would add nothing tot he model except complexity). The returned material is split into the two gas reservoirs, with a fraction $f_h$ going to the hot reservoir and the remainder becoming immediately available for star formation in the cold reservoir.
		
		The hot reservoir decays into the cold reservoir with a characteristic frequency $\lambda_\text{cool}$, but we also include a mechanism for stellar feedback: when stars of mass $m$ form from the cold gas, an additional amount $\delta m$ of cold gas is shifted into the hot reservoir. 
		
		We initially assume the the galaxy is composed only of cold gas ($M_c(t=0) = M_0$, $M_h(0) = M_*(0) = 0$). Subsequent infall from the IGM is parameterised by exponential laws:
		\begin{equation}
		\dot{M}_{c, \text{infall}}(t) = \sum_i M_i \beta_i \exp\left( -\beta_i t\right)
		\end{equation}
		Where the free parameters $\{M_i\}$ and $\{b_i\} = 1/\beta_i$ set the infalling mass and timescales respectively. Together, this produces the following coupled system of differential equations:	
		{\small
		\begin{align}
			\begin{split}
			\dot{M}_c & = \sum_i M_i \beta_i \exp\left( -\beta_i t\right) + (1-f_h) \mu M_* 
				\\
				& ~~~~~~~~~~~~ + \lambda_\text{cool}M_h - (1+\delta) \nu_\text{sfr} M_c
			\end{split}\label{E:McDot}
			\\
			\dot{M}_h & = f_h \mu M_* + \delta \nu_\text{sfr} M_c - \lambda_\text{cool}M_h
			\\
			{M}_\text{t} & \equiv M_c + M_h + M_*
			\\
			\dot{M}_\text{t} &= \sum_i M_i \beta_i \exp\left( -\beta_i t\right) ~\Longleftrightarrow ~M_\text{t} = \bar{M} - \sum_i M_i e^{-\beta_i t}\label{E:MtDot}
		\end{align}
		}
		This can be analytically solved for $M_c$, and hence the star formation rate $\rho_\text{SFR}(t) = \nu_\text{SFR} M_c(t)$. Because of the linearised assumptions we have made, the solution is expressible in terms of a sum of exponential terms.

	\subsection{Elemental Synthesis \& Return of Metals}

		We are chiefly interested in the chemical compostion of the cold gas reservoir at any given time, as this determines today's observed stellar surface abundances (with minor modifications due to i.e. dredge up or gravitational settling).
			
		If a nucleosynthesis pathway $j$ produces an amount $y_{j,X}(t)$ of element $X$ at time $t$, then the amount of $X$ due to $j$ present in the cold gas is given by $M_{xcj}$, the corresponding amount for the hot reservoir is given by $M_{xhj}$. They are linked via:
		\begin{align}
			\dot{M}_{xcj} & = (1-f_{h,j}) y_{j,X}(t) + \lambda_j M_{xhj} - (1+\delta) \frac{M_{xcj}}{M_{c}} \rho_\text{sfr}(t)
			\\
			\dot{M}_{xhj} & = f_{h,j} y_{j,X}(t) - \lambda_j M_{xhj} + \delta \frac{M_{xcj}}{M_{c}} \rho_\text{sfr}(t)
		\end{align}  
		The final terms in these equations arise due to star formation, which removes a fractional amount of the element from the cold gas reservoir and either heats it up through stellar feedback, or locks it up in stars. This simplifies to:
		\begin{align}
			\dot{M}_{xcj} & = (1-f_{h,j}) y_{j,X}(t) + \lambda_j M_{xhj} - (1+\delta) \nu_\text{sfr} F_\text{mod} M_{xcj} \label{E:ColdGasOrigin}
			\\
			\dot{M}_{xhj} & = f_{h,j} y_{j,X}(t) - \lambda_j M_{xhj} + \delta \nu_\text{sfr} F_\text{mod}  M_{xcj} \label{E:HotGasOrigin}
		\end{align}  	
		Here $F_\text{mod}$ has been introduced as a `lockup modification factor', such that the lockup rate is proportional to $F_\text{mod} \rho_\text{sfr}$, instead of just $\rho_\text{sfr}$. This modification is introduced to allow for the fact that $y$ is the rate at which new material is synthesised. Since stars are formed from polluted gas, as long as they do not destroy the material, they can release metals which they did not synthesise. If $F_\text{mod} < 1$, therefore, we reduce the rate at which material is being locked up by mimicking the recycling of previously synthesised material.  
	
	\subsection{Yield Functions}\label{A:Yields}
		
		To produce the yield functions, $y_{X,j}$, we invoke a Delay Time Distribution (DTD). This function, $\Psi_j(t)$ gives the probability of a star undergoing stellar death a time $t$ after it was formed\footnote{$\Psi_j(t) = \Psi_j(M,Z,t)$ is explicitly a function of the progenitor mass and metallicity in this formulation, but this is omitted from the notation for convenience}. The mass-rate of events $j$ (i.e. the stellar mass loss rate through channel $j$) occuring at a time $t$ is therefore given by:
		\begin{equation}
			R_j(t) = \int_0^t \rho_\text{SFR}(t - \tau) \Psi_j(\tau) \d \tau
		\end{equation}
		Swapping the integration variable $t^\prime = t - \tau$, it follows that the yield from event $j$ is given by:
		{\small
		\begin{align}
		\begin{split}
			y_{X,j}(t) = \int_0^\infty \zeta(M) \d M \int_0^t \rho_\text{SFR}(t^\prime) \Psi(t - t^\prime,M,Z_\text{cg}(\tau)) \times &
			\\
			Y_{X,j}(M,Z_\text{cg}(\tau)) &\d t^\prime 
		\label{E:GeneralYields}
		\end{split}
		\end{align}
		}	
		Where $\zeta(M)$ is the Initial Mass Function (IMF),$Y_{X,j}(M,Z)$ is the gross yield of $X$ from a star of mass $M$ and initial metallicty $Z$ dying through process $j$, and $Z_\text{cg}(t)$ is the cold-gas metallicity at a time $t$. With \eref{E:GeneralYields} in hand, we are able to derive three equations for the cases of CCSN events, collapsars and delayed/inspiral events such as SNIa and NSMas follows:

		\subsubsection{Core Collapse Supernovae}
			
			\eref{E:GeneralYields} simplifies for the case of CCSN from high-mass progenitors. Such CCSN occur at the end of a lifetime $T$, such that the DTD becomes a Dirac delta function: $\Psi(t) = \delta(t - T)$. In addition, for stars massive enough to go CCSN, this lifetime is short compared to the timescale over which galactic properties change, such that we may approximate $T\approx 0$:		
			\begin{equation}
				y_{X,\text{CCSN}}(t) = \rho_\text{SFR}(t) \int_0^\infty \zeta(M) Y(M,Z(t)) \d M \label{E:CCSNDependent}
			\end{equation}
			In the case where the yields are independent of the metallicty (which, by comparison with yield grids such as \cite{Chieffi2004,Limongi2018}, we see holds to a good approximaiton for both Fe and Mg), we may compute the integral over $M$ to find the characteristic yield function of $X$ via $j$, $\bar{Y}_{X,j}$, giving the synthesis rate via CCSN as:
			\begin{align}
			\begin{split}
				y_{X,\text{CCSN}}(t) & = \bar{Y}_{X,j} R_\text{CCSN}(t) 
				\\
				& = \bar{Y}_{X,j} \times \rho_\text{SFR}(t) \label{E:CCSNYield}
			\end{split}
			\end{align}

		\subsubsection{Collapsars}
			
			The yield function for collapsars follows largely the same reasoning as that presented above. However, we must account for the metallicity dependence collapsars are modelled to posses (see \S\ref{S:Cutoff}). Following the logic of \S\ref{S:MetalDecouple}, we decouple the background metallicity into an external function $Z_\text{cg} = \Z(t)$, and hence collapsar yields gain an additional factor $\tilde{a}(t,\tau,\Delta)$ such that:
			\begin{equation}
				\tilde{a}(t, \tau,\Delta) = \begin{cases} 1 ~~~~~ & t < \tau - \Delta \\ \frac{\tau - t}{\Delta} & \tau - \Delta \leq t < \tau \\ 0 & t \geq \tau \end{cases} \label{E:Suppressor}
			\end{equation}
			Combining this with \eref{E:CCSNYield}:	
			\begin{equation}
				y_{X,\text{Collapsar}}(t) = \bar{Y}_{X,\text{coll}} ~~\rho_\text{SFR}(t) ~~ \tilde{a}(t,\tc, \Delta) \label{E:CollYield}
			\end{equation}

		\subsubsection{Delayed Yields}
			
			Finally, we consider the yields of SNIa and Neutron Star Margers. These events do not occur uniquely at the end of a stellar lifetime. Instead, after the stellar lifetime has passed, there exists a period of probabilistic decay, whilst the system continues to evolve until finally the progenitors inspiral (for double degenerate SNIa and NSM events), or accrete enough matter from their companion (for single degenerate SNIa). In addition, there exists a non-trivial time delay before the first events can start occuring.
			
			Whilst the common DTD for SNIa is typically given as $\propto t^{-1}$, in order to continue our ability to easily analytically integrate them, we follow the work of \sbRef in using an exponential DTD for SNIa events - for a discussion of the validity of this approach, see \cite{Weinberg2017}. Hence:
			\begin{equation}
				\Psi_j(t) \propto \Theta(t - \tau_j) \exp\left( - \nu_j t\right) 
			\end{equation}
			Assuming metallicity independence, we find:
			\begin{align}
			\begin{split}
				y_{X,\text{delay}}(t,\nu,\tau) & = \bar{Y}_{X,j} \mathcal{I}\left[t,\rho_\text{SFR},\nu,\tau\right]
				\\
				& = \bar{Y}_{X,j} \Theta(t - \tau)\int_{\tau}^t \rho_\text{SFR}(t - x) \exp\left( - \nu x\right) \d x \label{E:DelayYield}
				\end{split}
			\end{align}
			The \cite{Iwamoto1999} W70 is a common metallicity-independent model for SNIa yields, though recent efforts such as have attempted to account for progenitor metallicity, \cite{Travaglio2005} for example shows that the Fe yields are to be altered by less than 6 per cent between 0.1$\Z_\odot$ and $Z_\odot$, such that we consider the metallicity-independent model a good approximation. 

\section{Simulation Model} \label{A:Simulation}
	
	\codename{} models the the galaxy as a series of concentric rings. Each annulus (of $\sim$0.2kpc in width) contains a number of gas reservoirs - each of which has an independent chemical composition and is assumed to be chemically well mixed - in addition to containing  stellar and stellar-remnant populations.
	
	As per the prescription of \sbRef (and unlike the analytical model in \ref{S:Model}) the composition of each annulus does not evolve independently: \codename{} incorporates both radial migration of stars due to resonant scattering (`churning') and the oscillation of stars around their guiding centres due to epicyclic motion (`blurring'), allowing for stars (and hence the chemicals they produce upon death) to migrate away from their place of birth. 
	
	In addition to the migration of stars, galaxies require a steady inflow of fresh gas in order to sustain sufficient star formation rates and avoid depletion \citep{Chiosi1980}, which must in turn drive radial flows of gas within the galaxy. We use the formalism of \cite{Bilitewski2012} to account for the angular momentum balance. The material accreted from the IGM is not pristine, but otherwise its composition is poorly constrained.  \cite{Schonrich2017} approximates the inflow composition using the abundance distribution of a ring in the mid-outer disc, but notes that this only materially affects the outer disc's metallicity.
	
	The chemical yields from exploding stars are taken from a number of sources in order to cover the wide range of mass and metallicity required: we produce a compiled grid from the data of \cite{Marigo2001,Chieffi2004,Maeder1992} and data retrieved from the ORFEO database of \cite{Limongi2008}. The stellar lifetimes as a function of mass and metallicity are extracted from the BaStI database of \cite{Pietrinferni2004}.

	\subsection{r-Process Yields}
		
		The r-process synthesis contributions from CCSN and Collapsars are added into the usual CCSN yield network. We have allowed $\bar{Y}_\text{CCSN}(M,Z) = \epsilon$, a small, constant level of synthesis to arise from CCSN throughout history. This parameter is calibrated to give a 2-5 per cent contribution at simulation end. We then add a collapsar contribution derived from \eref{E:FullYields}. Unlike \modelname{}, the Collapsar yields have the same thermal properties (i.e. distribution between hot and cold gas phase and cooling timescale) as standard CCSN gas, a consequence of the incorporation into the standard yield network. 
		
		NSM contributions are modelled as in SW19 - they are treated near-identically to SNIa events, with the exception that they have a shorter initial delay time (their projenitors are higher mass objects, and so have shorter lifetimes), and a lower hot-gas injection fraction - that is, a larger fraction of gas from NSM events is immediately available for star formation than CCSN or SNIa events, as justified in \S\ref{S:Phasing}.
		
		As with \modelname{} each of these pathways -- NSM events, collapsars and small s-process contribution --  has an undetermined prefactor, these are chosen either to reproduce the observational data, or to target some ideal collapsar/NSM/s-process fraction at simulation end (or some combination of both), though due to the numerical complexity of the simulation, the values have to be tuned by hand, rather than using simple analytical constraints.

\bsp	
\label{lastpage}
\end{document}